\documentclass[referee]{mn2e}

\usepackage{graphicx}
\usepackage{amsmath}
\usepackage{url}

\title[High energy radiation from magnetized stars]
{High energy radiation from luminous and magnetized stars}
\author[W. Bednarek]
{W. Bednarek\\ 
University of \L \'od\'z, Department of Astrophysics, Faculty of Physics and Applied Informatics,
ul. Pomorska 149/153, 90-236 \L \'od\'z, Poland,\\
wlodzimierz.bednarek@uni.lodz.pl\\}
\begin{document}

\date{Accepted . Received ; in original form }

\pagerange{\pageref{firstpage}--\pageref{lastpage}} \pubyear{2015}

\maketitle

\label{firstpage}

\begin{abstract}
A part of early type stars is characterised by strong dipole magnetic field that is 
modified by the outflow of dense wind from the stellar surface. At some distance from the surface 
(above the Alfven radius), 
the wind drives the magnetic field into the reconnection in the equatorial region of 
the dipole magnetic field. We propose that  electrons accelerated in these reconnection 
regions can be responsible for efficient comptonization of stellar radiation producing 
gamma-ray emission. We investigate the propagation of electrons in 
the equatorial region of the magnetosphere by including their advection with 
the equatorial wind. The synchrotron and IC spectra are calculated assuming
that a significant part of the wind energy is transferred to relativistic electrons. 
As an example, the parameters of luminous, strongly magnetized star HD 37022 ($\Theta^1$ Ori C) are considered. The IC gamma-ray emission is predicted to be detected either 
in the GeV energy range by the Fermi-LAT telescope or in the sub-TeV energies by the 
Cherenkov Telescope Array. 
However, since the stellar winds are often time variable and the magnetic axis can be inclined 
to the rotational axis of the star, the gamma-ray emission is expected to 
show variability with the rotational period of the star and, on a longer time scale, with 
the stellar circle of the magnetic activity. Those features might serve as tests of the proposed
scenario for gamma-ray emission from single, luminous stars. 
\end{abstract}

\section{Introduction}

About ten percent of luminous massive stars (OB type) show strong, dipole type 
magnetic fields (Stahl et al.~1996, Donati et al.~2002), 
that  influence the structure of fast winds from 
stellar surfaces. A part of the wind, launched in the equatorial region of the star, is focused 
towards the magnetic equator. Two streams of plasma from opposite hemispheres collide in 
the equatorial plane heating the medium to temperatures of the order of several $10^6$ K.
(Shore \& Brown~1990, Babel \& Montmerle~1997, Gagne et al.~2005).
On the other hand, the wind launched closer to the polar regions can propagate
from the star drawing the magnetic field out. As a result, the winds from massive stars are 
highly anisotropic as envisaged by e.g. Shore~(1987).
The polar wind is focused by the magnetic field into 
the neutral sheet in the equatorial region of the magnetic dipole. It is argued that 
in this region, electrons are accelerated in the process of magnetic reconnection 
(e.g. Usov \& Melrose 1992). Those electrons are expected to produce non-thermal X-ray radiation. 
In fact, strong X-ray emission
is detected from early type single massive stars (e.g. Pollock 1987, Chlebowski 1989).
Relativistic electrons are also expected to produce non-thermal radio emission, observed 
in the case of 25\% of  massive stars (e.g.  Linsky et al.~1992). 
The model for the acceleration of electrons in the reconnection regions of the equatorial
part of the wind in the outer magneto-sphere of a massive star has been more recently applied 
to explain its non-thermal
radio and X-ray emission (see e.g. Trigilio et al.~2004, Leto et al.~2006, Leto et al.~2017). 
In this model, electrons are assumed to be accelerated in the reconnection regions in the outer 
magnetosphere, as postulated by Usov \& Melrose (1992). They follow magnetic 
field lines towards the stellar surface producing synchrotron radiation in the polar regions of 
the star as observed in the case of Jupiter (Branduardi-Raymont 2007). 

Here we apply the general scenario for the acceleration of electrons in the reconnection regions
of the outer stellar magnetosphere mentioned above. However, we argue that electrons can reach 
multi-TeV energies in the reconnection regions. After injection from the reconnection regions, 
they are isotropised in the random component of the magnetic field present 
in the equatorial wind. In contrast to Trigilio et al.~(2004), we assume that
electrons are advected with the equatorial part of the wind in the outward 
direction from the star. In such general scenario, we consider radiation processes, the synchrotron 
and the Inverse Compton, 
which turn to the production of $\gamma$-rays.  We investigate the conditions for which the 
high energy radiation, produced in the vicinity of isolated massive star, can be detectable by the
present and future $\gamma$-ray observatories. As an example, the case of the 
nearby, luminous, strongly magnetized star HD 37022 ($\theta^1$ Ori C) is considered. We show that the parameters of this star allow acceleration of electrons up to TeV energies.

In fact, production of $\gamma$-rays in the ICS process of relativistic electrons accelerated in 
the stellar wind has been already considered from the beginning of the $\gamma$-ray astronomy 
(e.g. White 1985, Pollock 1987). 
However, another model for the acceleration of electrons in the winds of massive stars has 
been applied. In the multiple shock in the stellar wind model 
(originally introduced by Lucy \& White 1980 and Lucy 1982), the acceleration of electrons (and 
hadrons) occurs in the non-stationary wind in which shocks are formed 
(Chen \& White~1991, White \& Chen~1992). However, magneto-hydro-dynamic (MHD) simulations of 
the propagation of the wind do not support formation 
of such efficient shocks close to the star (e.g. Owocki \& Rybicki 1984). 

Stars (including the Sun) have been also more recently predicted to be steady sources of 
$\gamma$-rays produced by cosmic ray electrons which IC up-scatter stellar radiation (Moskalenko et al.~2006, Orlando \& Strong 2007,2008,2021). In fact, such $\gamma$-ray emission has been detected by the EGRET and the Fermi-LAT telescopes (Orlando \& Strong 2007, Abdo et al. 2011). The $\gamma$-ray emission from the IC process from the nearby super-luminous stars has been also predicted to be within the sensitivity limits of the Fermi-LAT telescope. However, up to now the search of the 12 years of Fermi-LAT data from 9 super-luminous stars in the solar vicinity resulted only in the upper limits which allows to constrain cosmic ray electron spectra in the vicinity of those isolated stars (de Menezes et al. 2021). Luminous stars in dense clusters seems to be stronger emitters of $\gamma$-rays, in terms of discussed above scenario, since they are exposed to much stronger density of relativistic electrons accelerated in a turbulent environment (e.g. Ackermann et al. 2011).

\section{Model for the stellar wind/magnetic field interaction}

As mentioned above, we apply the general model for the structure of the winds around magnetized 
luminous stars. It is assumed that the star has ordered (dipole type) structure of the magnetic
field.  
The luminous star produces strong stellar wind which interacts with a strong dipole 
type magnetic field around the star. We assume that at the considered range of distances from 
the star, the wind has the velocity which already does not differ significantly from the tangent velocity at the infinity.
The wind is launched isotropically from the stellar surface with constant density.
As a result of the stellar wind/magnetic field interaction, 
two regions can be defined in the stellar magnetosphere  (Babel \& Montmerle~1997). 
Below the so called Alfven radius, i.e. in the closed magnetosphere, magnetic field drives 
the wind into the equatorial plane of the magnetic dipole producing two streams of oppositely 
propagating plasma (see Fig.~1a). The collision of those two streams forms very hot region which emits soft, 
thermal X-ray emission. This is so called closed magnetosphere.
Above the Alfven radius, the wind drives the magnetic field towards the equatorial region forming 
the neutral sheet in which reconnection process of the magnetic field 
can become efficient. We argue that the electric field, induced in the reconnection region, is able to  accelerate electrons even to multi-TeV energies.

\begin{figure*}
\vskip 7.truecm
\includegraphics{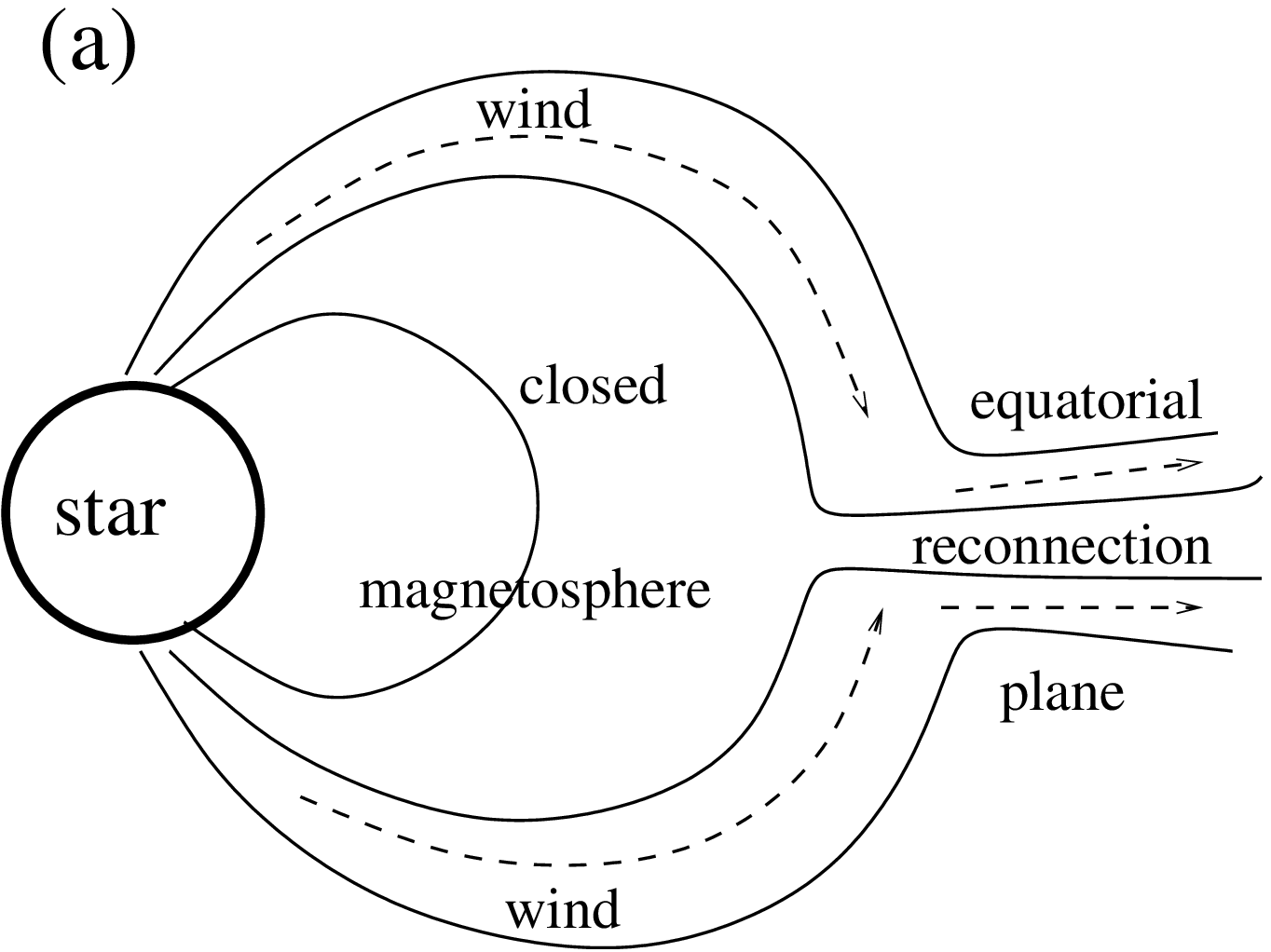}
\includegraphics{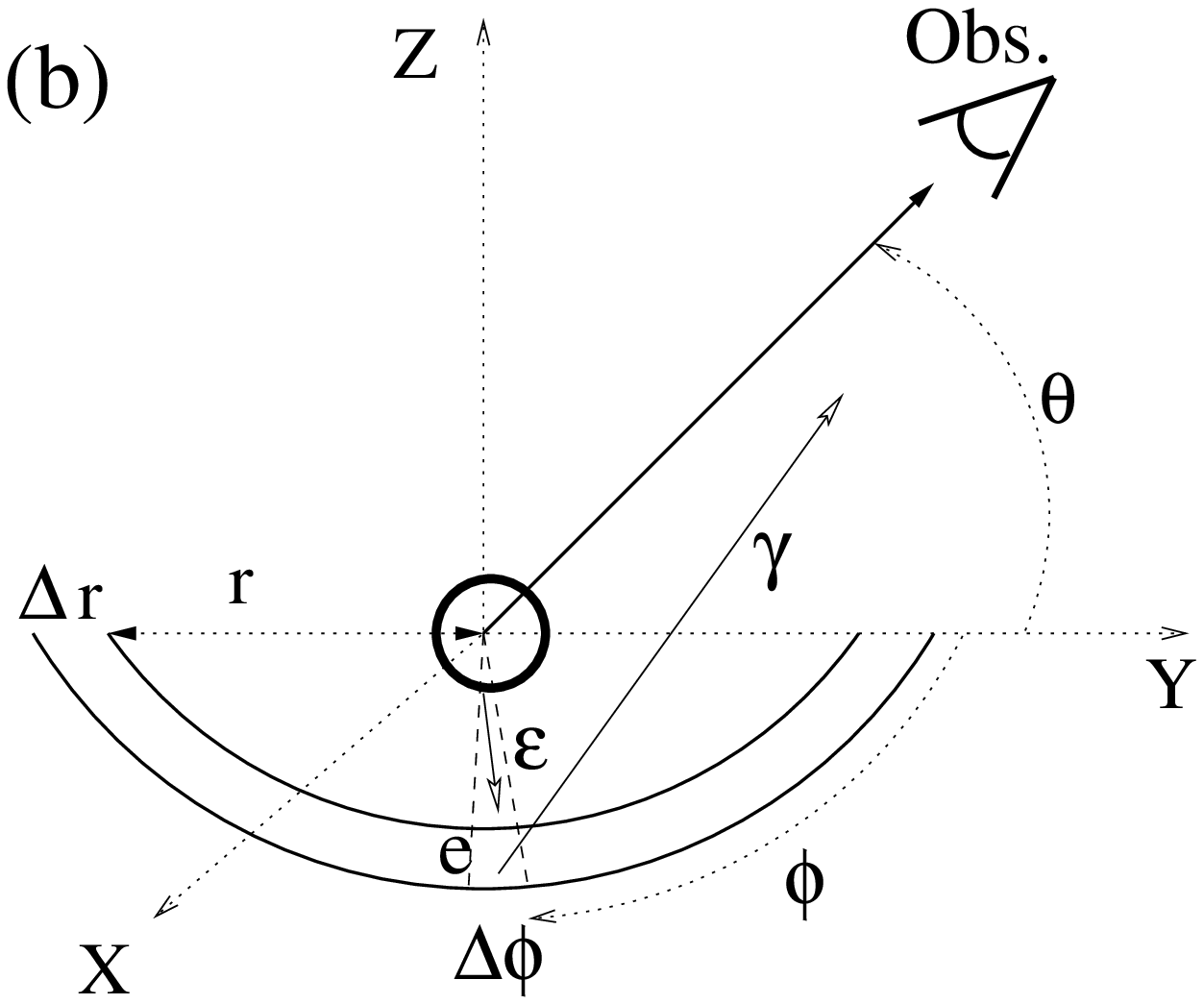}
\caption{Figure (a): Schematic presentation of the considered model. A luminous star 
produces a dipole magnetic field and launches a strong wind. Close to the star, below the Alfven 
radius, the wind flows along the magnetic field lines. Since the energy density of the magnetic 
field drops faster with the distance from the star than the energy density of the wind, 
the wind starts to dominate 
the flow above the Alfven radius. Then, the magnetic field is forced into reconnection in the 
equatorial plane. Electrons are
accelerated in the reconnection region to maximum energies depending on the velocity of the wind 
and the magnetic field strength. They are injected with the power law spectrum into the turbulent 
equatorial region. Electrons lose energy on the synchrotron process (in the
random magnetic field) and on the Inverse Compton process (by scattering stellar radiation). 
Electrons are advected along the equatorial plane with the velocity of the equatorial wind.
Figure (b): The geometry of the IC scattering process of stellar radiation ($\varepsilon$) by 
an electron (e) within the narrow ring of the equatorial disk with the radius, r, the 
thickness $\Delta$r, and at the azimuthal angle 
$\phi$ (within the range $\Delta \phi$). Electron (e) produces $\gamma$-ray photon ($\gamma$) 
propagating towards the observer located at the angle, $\theta$, in respect to the equatorial 
disk surface.}
\label{fig1}
\end{figure*}

We consider the example star which is characterised by the mass loss rate as seen at the infinity, 
$\dot{M}_{\rm w} = 4\pi R^2\rho v_{\rm w}$,
where the distance from the star is $R = r R_\star$ with the stellar radius 
$R_\star = 10^{12}R_{12}$~cm, the velocity of the wind is $v_{\rm w} = 10^8v_8$ cm s$^{-1}$, and $\rho$ 
is the wind density. The star has dominant dipole magnetic field structure with the strength,
$B(r) = 0.5B_\star r^{-3}$, where $B_\star = 10^3B_3$ G is the surface magnetic field strength at 
the magnetic pole.
By comparing the energy density of the wind with the energy density of the magnetic field, 
we estimate the distance (the Alfven radius) from the star at which the wind starts to drive 
frozen in magnetic field.  This distance is at, 
\begin{eqnarray}
r\approx 3.4 (B_3R_{12})^{1/2}/(\dot{M}_{-6}v_8)^{1/4}, 
\label{eq1}
\end{eqnarray}
\noindent
where
the mass loss rate of the wind is $\dot{M}_{\rm w} = 10^{-6}\dot{M}_{-6}$~M$_\odot$ yr$^{-1}$.
Above this distance, the magnetic field, carried by the wind, can efficiently 
reconnect in the equatorial region of the dipole magnetic field (Usov \& Melrose~1992, Trigilio et al.~2004, Leto et al.~2017).
At this region, the wind propagates in the outward direction of the star. It can continue flowing
in the outward 
direction forming the equatorial part of the wind as often envisaged in MHD simulations
(e.g. ud-Doula \& Owocki~2002). 
Since the stellar winds are expected 
to be turbulent, electrons injected from the reconnection region are likely isotropised 
in the wind reference frame.

The maximum energies of electrons accelerated in the reconnection region can be estimated as a 
product of the induced electric field and the  length scale of the reconnection region.
The electric field can be parametrised as, 
$E_{\rm rec} = \varepsilon \beta_{\rm w}cB$, where $\beta_{\rm w} = v_{\rm w}/c = 0.01\beta_{0.01}$ 
is the relative velocity of the wind in respect to the velocity of light $c$, and 
$\varepsilon = 0.1\varepsilon_{0.1}\sim 0.1$ is the reconnection efficiency, 
e.g. Uzdensky~(2007). The length scale of the reconnection region can be parametrised by, 
$L_{\rm rec} = \xi R_\star r$ cm, where $\xi$ is the coefficient describing the 
extend of the reconnection region defined as a  part of distance of the reconnection 
region from the centre of the star. Then, the maximum energies of electrons are, 
\begin{eqnarray}
E_{\rm max} = E_{\rm rec}L_{\rm rec} = 3\varepsilon_{0.1} \xi \beta_{0.01} 
B_3R_{12}r_{10}^{-2}~~~{\rm TeV},
\label{eq2}
\end{eqnarray}
\noindent
where $r = 10r_{10}$ is the distance from the star in units of the stellar radii.
We assume that electrons escape with the power law spectrum from the reconnection region into 
the equatorial wind. The spectrum is characterised by the exponential cut-off,
\begin{eqnarray}
{{dN}\over{dE}} = A E^{-\alpha}exp(-E/E_{\rm max}),
\label{eq3}
\end{eqnarray}
\noindent
where $A$ is the normalization constant, $E_{\rm max}$ is estimated from Eq.~2, and 
$\alpha$ is the spectral index.

In summary, in the considered scenario, electrons are accelerated in the equatorial disk which 
extends between the range of distances $R_{\rm min}$ and $R_{\rm max}$ from the star. 
After isotropisation in the turbulent wind, they are advected 
in the outward direction from the star. Electrons lose energy on the synchrotron and the IC processes. 
Electrons, at specific distance, $R$, from the centre of the star, comptonize stellar radiation 
towards the observer 
located at the angle $\theta$ to the disk plane (see Fig.~1b). The effectiveness of the $\gamma$-ray
production depends on the location of electrons within the disk (defined by the distance, $R$, and 
the azimuthal angle, $\phi$). $\gamma$-rays, produced in the IC process, are partially re-absorbed in the stellar radiation. $\gamma$-ray spectra are investigated for different 
parameters of the acceleration model and the location of the observer in respect to the equatorial disk.

\section{Propagation of electrons in the equatorial wind}

Electrons, injected from the reconnection regions, are confined in the plane of the equatorial 
wind since their Larmor radii, in the local magnetic field, are very small in respect to the typical 
distance from the star, i.e. 
\begin{eqnarray}
R_{\rm L} = E_{\rm max}/(eB)\approx 10^{10}\varepsilon_{0.1} \xi 
\beta_{0.01}R_{12}r_{10}~{\rm cm}<<{\rm R}, 
\label{eq4}
\end{eqnarray}
\noindent
where $e$ is the charge of the electron. Those electrons are advected with the equatorial wind 
in the outward direction from the star with the constant wind velocity.
Electrons suffer energy losses on the Inverse Compton Scattering of stellar 
radiation and on the synchrotron process. The relative role of those two energy loss processes
(i.e. in the case of IC in the Thomson regime) can be easily estimated by comparing the energy 
densities of the stellar photons and the magnetic field 
as a function of distance from the star.
The energy density of stellar photons is given by $\rho_{\rm rad}\approx 90 T_4^4r^{-2}$ 
erg~cm$^{-3}$, assuming that the stellar radiation is well approximated by the black body 
spectrum with temperature  $T = 10^4T_4$~K. 
The energy density of the magnetic field is $\rho_{\rm B}\approx 4\times 10^4B_3^2r^{-6}$ 
erg~cm$^{-3}$. From the comparison 
of $\rho_{\rm rad}$ and $\rho_{\rm B}$, we estimate the distance, 
\begin{eqnarray}
r\approx 4.6 B_3^{1/2}/T_4,
\label{eq5}
\end{eqnarray}
\noindent 
above which the IC losses (in the Thomson regime) dominate over the synchrotron energy losses. 
For typical parameters, this distance is larger than the Alfven radius (Eq.~1). This means that the
dominant energy loss process of electrons is synchrotron radiation in the inner part of 
the equatorial wind. The IC process (in the Thomson regime) dominates at farther distances from 
the star.
Note however, that in the Klein-Nishina regime, i.e. for electrons with energies above 
$E_{\rm e}\sim 100/T_4$ GeV, the synchrotron process becomes important even at large distances 
from the star.

We also estimate of the optical depth on the Inverse Compton scattering of 
stellar radiation by relativistic electrons in order to have impression how efficiently
energy is transferred from relativistic electrons to the high energy radiation. 
The optical depth can be estimated from,
$\tau = n_{\rm ph} \sigma_{\rm T} R\approx 1.3T_4^3R_{12}r_{10}^{-1}$,
where $n_{\rm ph}\approx 1.9\times 10^{11}T_4^3r_{10}^{-2}$~cm$^{-3}$ is the density of 
stellar photons. We conclude that a significant amount of electron's initial energy 
is transferred to the high energy $\gamma$-rays in 
the IC process, provided that the surface temperature of the luminous star is clearly above 
$10^4$~K.

In order to follow the advection process of electrons, we divide the equatorial 
disk on a sequence of rings with the radius $r$ and the thickness $0.2r_\star$. 
Electrons are assumed to be injected at a specific element of the ring defined by the azimuthal 
angle $\phi$ (measured from the location of the observer, see Fig.~1b).

\section{Radiation from electrons}

In order to calculate the $\gamma$-ray spectra, produced by electrons in the comptonization of stellar 
radiation, we modify the numerical Monte Carlo code for the development of the IC $e^\pm$ pair 
cascade initiated 
by relativistic electrons in the anisotropic radiation from the luminous star, originally developed 
by Bednarek~(1997,2006). The code calculates the $\gamma$-ray spectra from
electrons located at a specific distance from the stellar surface. $\gamma$-rays escape 
towards the observer located 
at an arbitrary angle to the direction between the electron and the centre of the star.
It takes into account the Inverse Compton process and the $e^\pm$ pair production by $\gamma$-rays
in collision with stellar radiation. In the present calculations, we limit the  IC $e^\pm$ pair 
cascade process to the first generation of produced $\gamma$-rays in the IC process since 
it becomes very complicated to follow
emission from secondary $e^\pm$ pairs in complex structure of the magnetic
field influenced by the out-flowing stellar wind in the present scenario. The structure of 
the magnetic field can strongly influence the paths of secondary cascade $e^\pm$ pairs, 
re-distributing directions of secondary $\gamma$-rays (e.g. Sierpowska \& Bednarek~2005). 
Therefore, the $\gamma$-ray spectra, obtained in our scenario, should be considered as 
lower limits, especially at their low energy part. 
Due to the anisotropic radiation field of the star (as seen from the location of relativistic 
electrons), the $\gamma$-ray spectra from the IC process depend on the inclination angle of 
the observer in respect to the equatorial disk. 

The  synchrotron spectra are obtained assuming dominant random magnetic field in the equatorial 
disk outside smaller scale coherent reconnection regions. The cooling process of electrons, 
injected with energy $E_{\rm e}(r)$, is followed up to the moment of their advection
from the disk or their cooling to the energy $E_{\rm min} = 1$~GeV. Electrons with 
$E_{\rm min}$ are not able to produce $\gamma$-rays above 100 MeV in the IC process. 
Due to a relatively small velocity of the equatorial 
wind (beaming of radiation not essential) and the assumption on the random magnetic field within 
the disk, the synchrotron emission is isotropic.
 
As we mentioned above, relativistic electrons are injected
within the thin disk laying in the equatorial plane of the dipole magnetic field.
Electrons have a power law 
spectrum defined by the spectral index, the maximum energy determined by the  
conditions within the disk, and the power corresponding to  ten percent of the power of 
the stellar wind (Usov \& Melrose 1992). This normalization to the wind power bases 
on the assumption that the wind is launched isotropically 
from the stellar surface. We estimate a part of this isotropic wind which falls onto the 'n'th ring
extending between distances $R_{\rm n}$ and $R_{\rm (n+1)}$ (see also Trigilio et al.~2004).
The thickness of the ring has been chosen on $\Delta R = 0.2r_\star$, 
i.e. $R_{\rm (n+1)} = R_{\rm n} + 0.2R_\star$.
The equation of a specific magnetic field line, with the dipole structure, is given by 
$R_\star = R_{\rm n} \sin^2{\lambda_{\rm n}}$, 
where $\lambda_{\rm n}$ is the angle between the magnetic pole and the starting point of 
the magnetic field line extending to $R_{\rm n}$, and $R_{\rm n}$ is the distance measured 
in the equatorial plane of the magnetic 
dipole. We want to derive a part of the solid angle between the magnetic 
field lines starting at the angles $\lambda_{\rm n}$ and $\lambda_{\rm (n+1)}$. 
They correspond to the extend of the ring. This part of the whole solid angle is  
$\Delta\Omega/(4\pi) = 0.5(\cos{\lambda_{\rm n}} - \cos{\lambda_{\rm (n+1)}})$. Only this part 
of the wind, launched from the star, falls onto the ring within the radii 
$R_{\rm n}$ and $R_{\rm (n+1)}$.

Electrons are quickly isotropised by a turbulent magnetic field within the equatorial
disk. They are assumed to lose energy on two processes. The synchrotron process in the local
magnetic field and the IC process in the anisotropic radiation field of the stellar radiation.
We assume that electrons lose energy continuously on the synchrotron process up to the moment of 
the first production of $\gamma$-ray photon. The energy and emission angle of produced 
$\gamma$-ray photon is counted in order to produce angular dependent $\gamma$-ray spectra in respect 
to the equatorial disk.  In this calculations, we also take into account the change of the 
position 
of the electron on the equatorial disk surface due to their advection process outside the equatorial 
wind. The cooling process of electrons and the process of $\gamma$-ray production is 
followed by applying the Monte Carlo method. We calculate the $\gamma$-ray 
spectra from  every element of the ring. The location of this surface element is defined by 
the distance from the star, $R$, and the azimuthal angle, $\phi$, in respect to the location of the 
observer, whose position is defined by the inclination angle, $\theta$, to the plane of the 
equatorial disk.

\section{Application to HD 37022 ($\theta^1$ Ori C)}

We apply the model for the example luminous, magnetized O type star. We select the parameters of the star in the way that the acceleration of electrons in the reconnection region is possible to TeV energies (see Eq. 2). Moreover, the radiation field of this example star has to guarranttee efficient IC scattering of stellar radiation by those electrons (i.e. optical depth above unity, see Sect.~3). As noted in the introduction,
significant part of massive stars show strong magnetic field. Detailed structure and strenght of the field is precisely measured only in the case of several O type stars (see Table~1 in Chandra et al. 2015 and Petit et al.~2013).  
The best parameters, from the modelling point of view, has the star  HD 37022, 
which is also known as $\Theta^1$ Ori C. 
From another side, $\Theta^1$ Ori C is a problematic target for $\gamma$-ray telescopes since
it is the most luminous star in the Trapezium cluster which is a part of the Orion Molecular Complex. The Orion Complex is well extablished very extended $\gamma$-ray source
(Ackermann et al.~2012a,2012b).
Therefore, investigation of a point like $\gamma$-ray source in front of the extended 
$\gamma$-ray source can become problematic for the analysers of the Fermi-LAT data.
The basic parameters of the main star 
in $\Theta^1$ Ori C, such as the radius $R_\star$, the surface temperature $T_\star$  
(Sim\'on-Diaz et al.~2006), the dipole magnetic 
field strength at the pole $B_\star$ (Wade et al.~2006),  
the wind velocity $v_{\rm w}$, and the mass loss rate $\dot{M}_{\rm w}$ (Petit et al.~2013), 
are given in Tab.~1. For the above parameters, the wind power is
$L_{\rm w} = \dot{M}_{\rm w} v_{\rm w}^2/2\approx 1.56\times 10^{36}$~erg~s$^{-1}$.
The distance to HD 37022 is equal to 410 pc (Kraus et al.~2009).
In fact, HD 37022 is a binary system in which less luminous B type
star has a relatively low mass loss rate. Separation of the companion stars at 
the periastron passage is $(1-e)a\approx 2.8\times 10^{14}$ cm,  
for the semi-major axis of the binary system $a = 40$~AU and the eccentricity $e=0.534$ 
(Kraus et al.~2009).  This is 
clearly larger than the distance scale of the colliding wind  around HD 37022. 
Therefore, we conclude that the presence of the companion star in this binary system does not 
influence processes in the inner magnetosphere around more luminous companion HD 37022.

For the above parameters, the closed magnetosphere around HD 37022 extends only to $2.4R_\star$. 
Above this distance, the magnetized wind of HD 37022 forms good conditions for efficient 
reconnection of the stellar magnetic field. In these reconnection regions, electrons are 
expected to be accelerated to  multi-TeV energies. After initial acceleration 
process in the reconnection regions, electrons are slowly advected along the equatorial 
wind region in the outward direction from the star. They lose energy on the synchrotron 
radiation and the inverse Compton up-scattering of thermal radiation from the nearby surface 
of the luminous star.

\begin{table}
  \caption{Parameters of the star HD 37022 ($\theta^1$ Ori C)}
  \begin{tabular}{lllll} 
\hline 
\hline 
 R$_\star$ (cm)  & T$_\star$ (K)  &  B$_\star$ (G) & $\dot{M}$ (M$_\odot$/yr) &  v$_{\rm w}$ 
(cm/s) \\
\hline
$7\times 10^{11}$  & $3.9\times 10^4$ & $1.1\times 10^3$ & $5\times 10^{-7}$ & $3.225\times 
10^8$ \\
\hline 
\hline 
\end{tabular}
  \label{tab1}
\end{table}

\begin{figure*}
\vskip 8.5truecm
\includegraphics{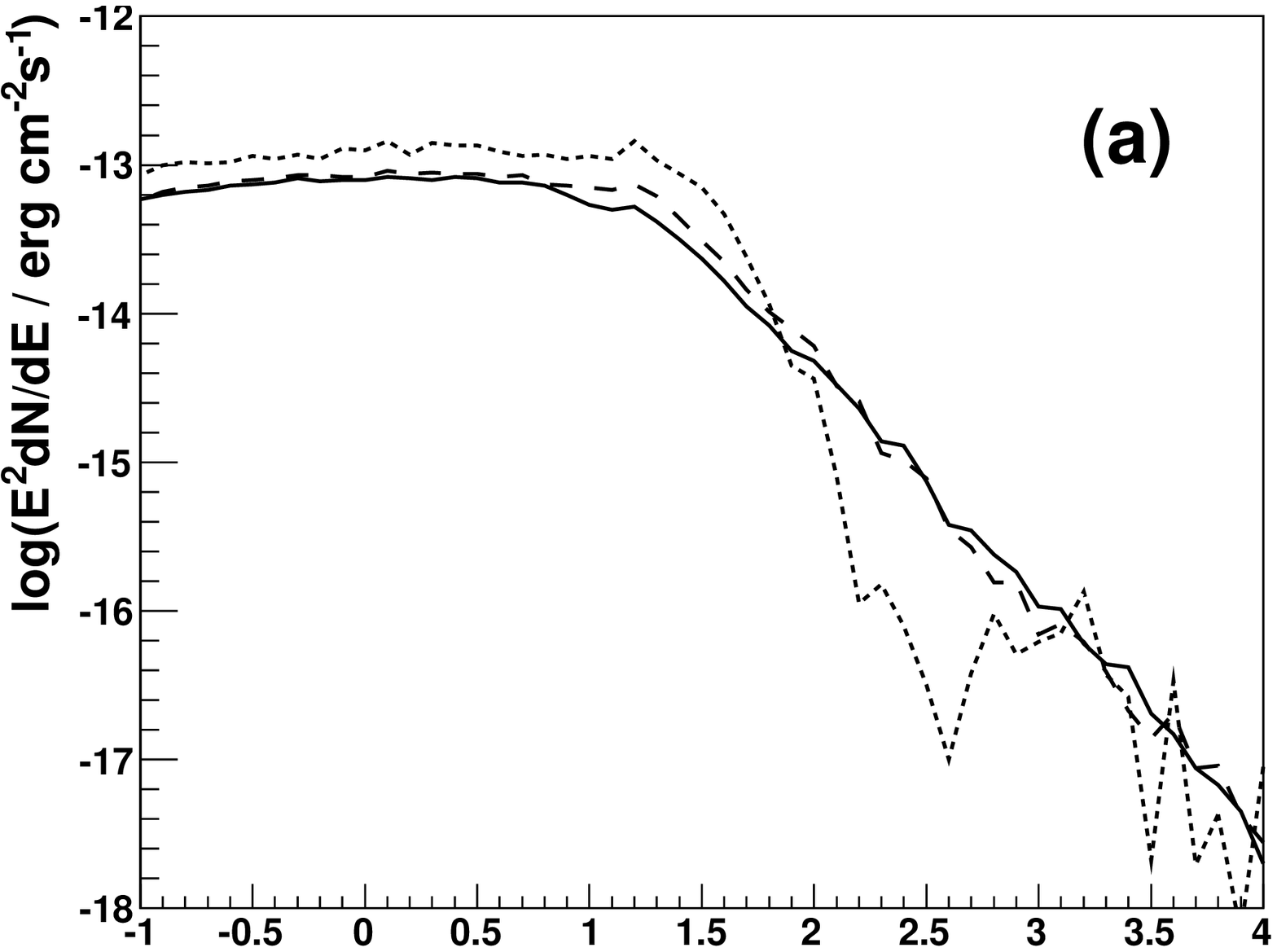}
\includegraphics{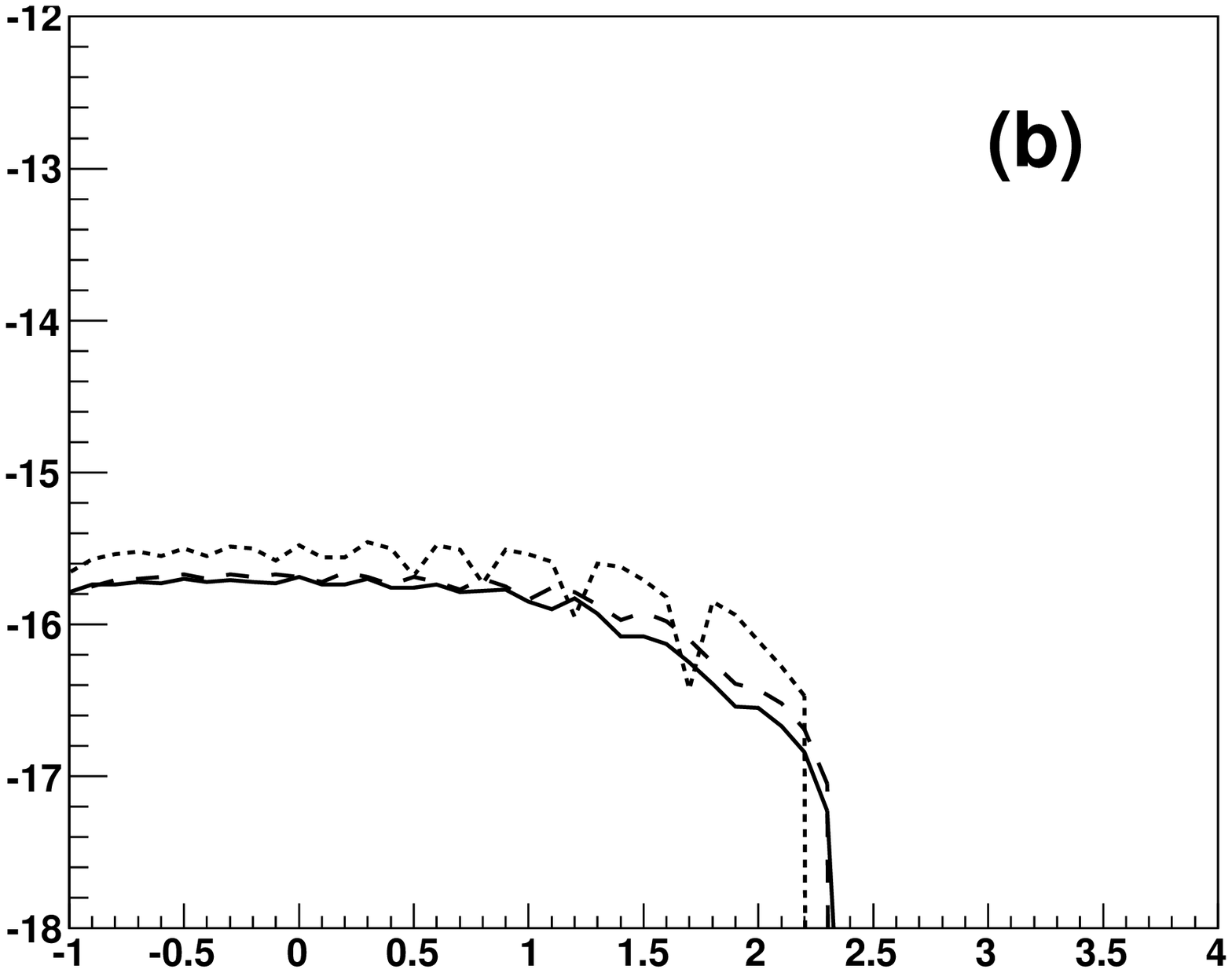}
\includegraphics{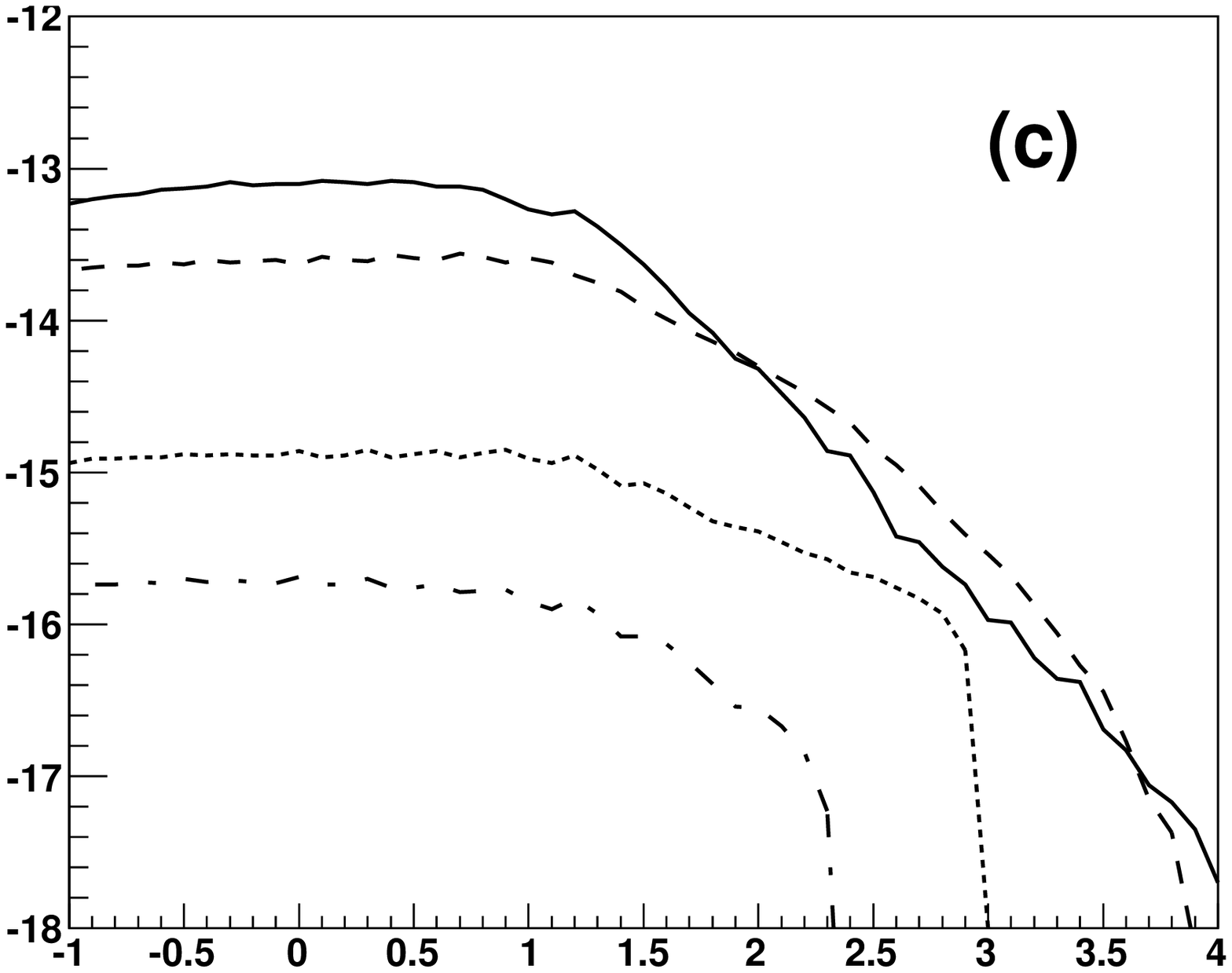}
\includegraphics{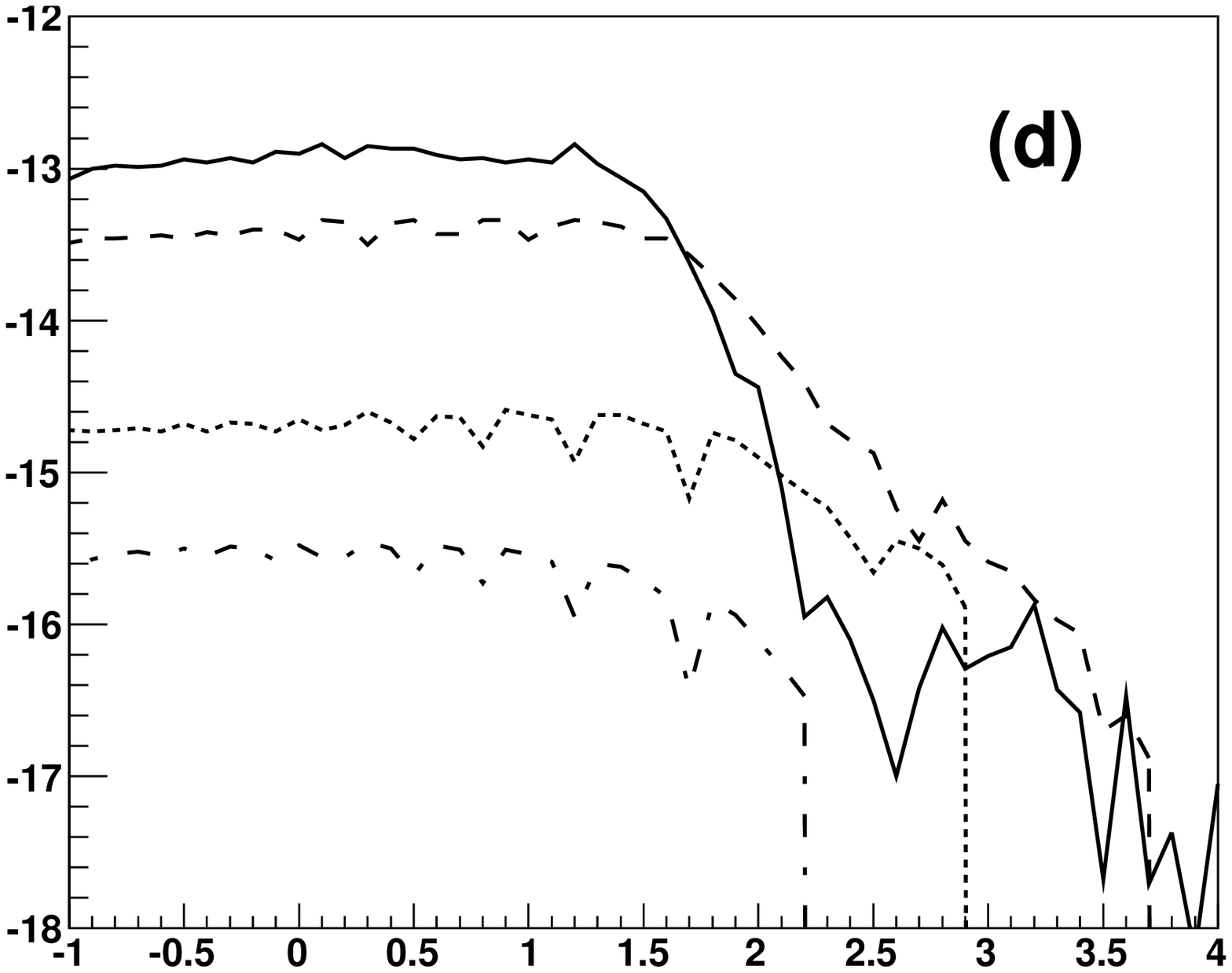}
\includegraphics{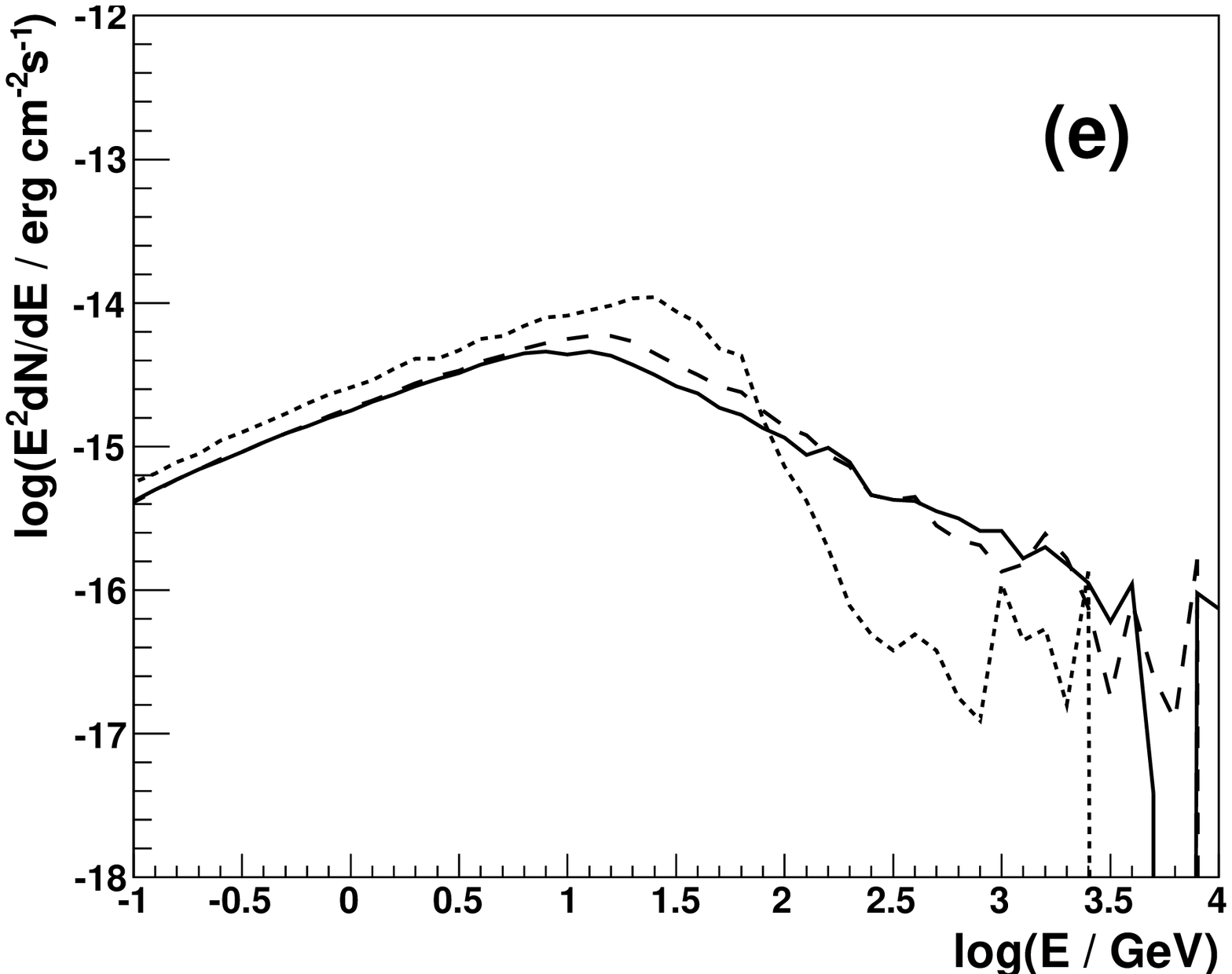}
\includegraphics{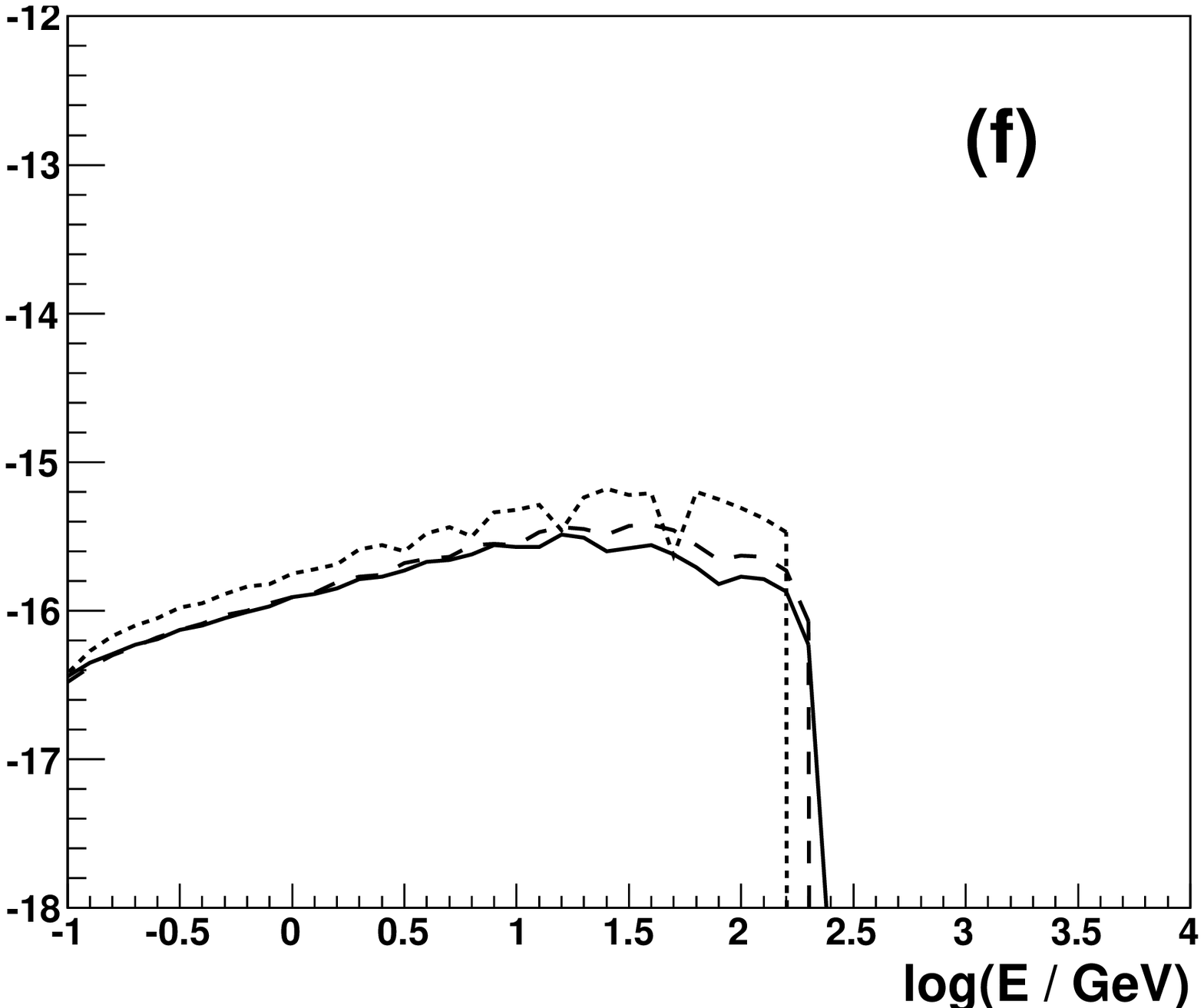}
\includegraphics{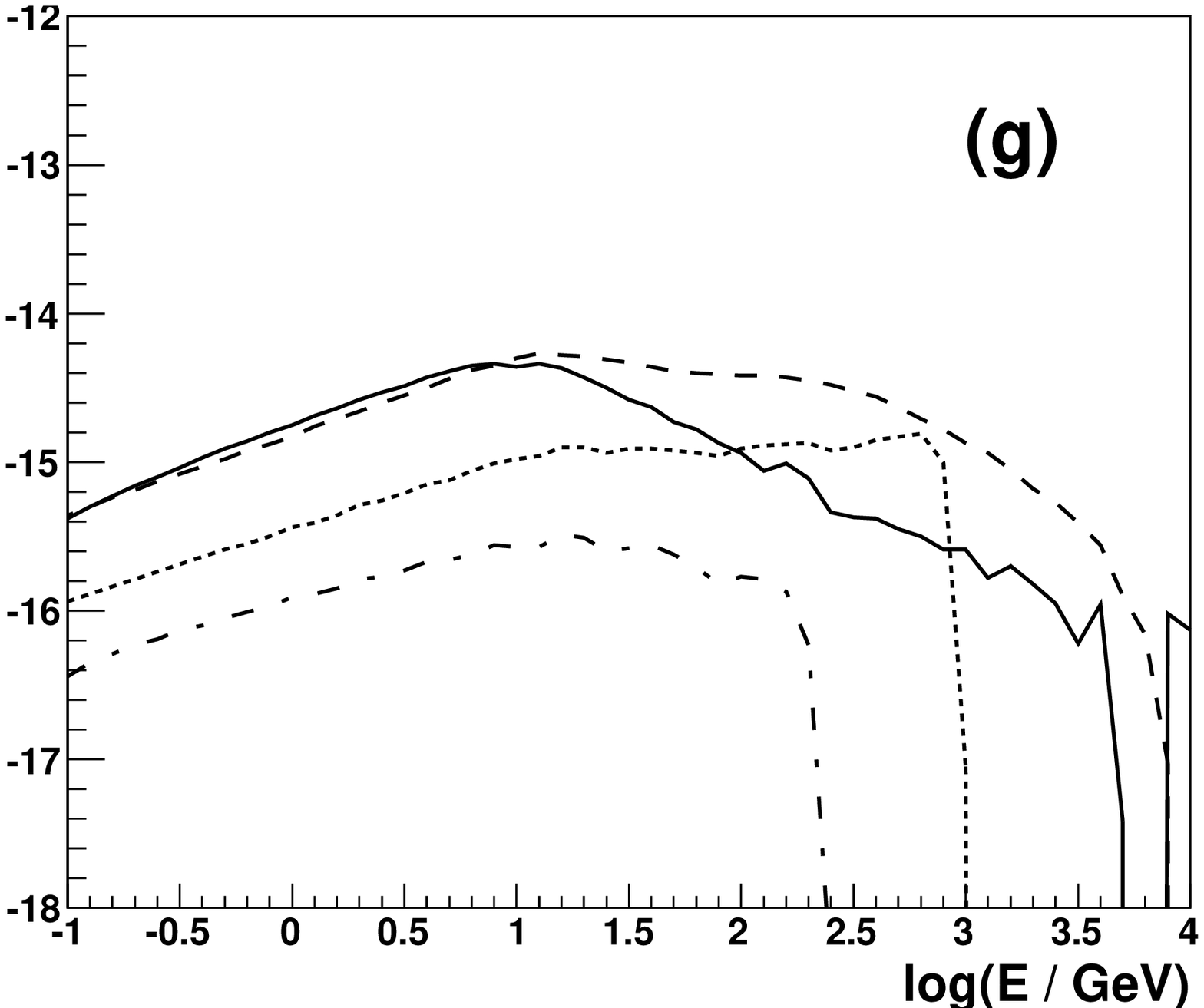}
\includegraphics{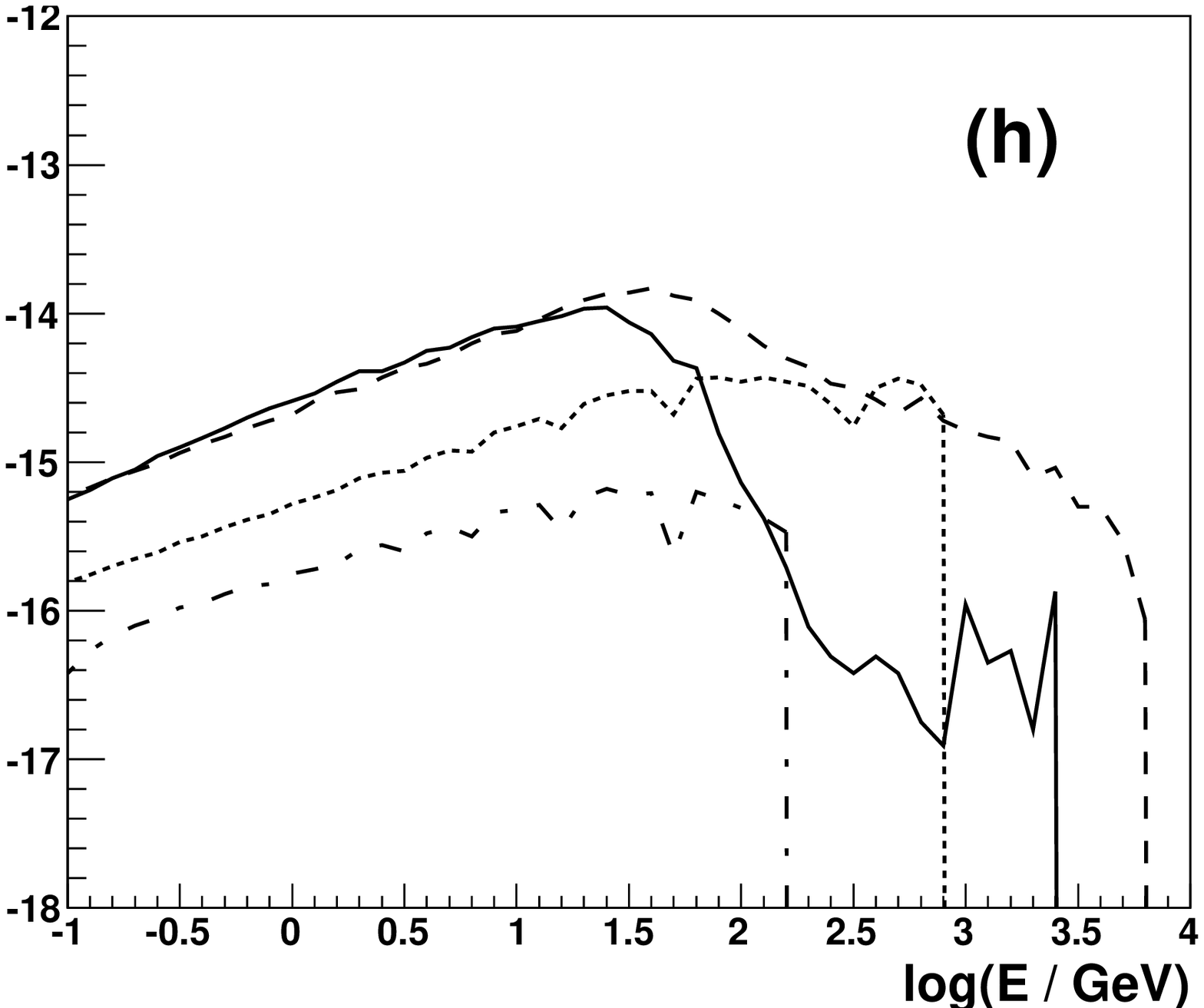}
\caption{Gamma-ray spectra produced by relativistic electrons in the IC scattering process 
of thermal radiation from the star with the example parameters of HD 37022 (see Table~1). 
Electrons are accelerated in the reconnection region extending in the plane perpendicular 
to the axis of the dipole magnetic field of the star. Electrons are injected 
from the reconnection regions in the equatorial wind region with  the power spectrum characterised 
by the spectral index, $\alpha$, up to the maximum energy, $E_{\rm max}$, defined by Eq.~2.
Electrons are injected at  specific distances from the surface of the star: (panel a) 
$R = 2.5R_\star$ 
and $\alpha = 2$; (e) $R = 2.5R_\star$ and $\alpha = 1$; (b) $R = 30R_\star$ and $\alpha = 2$;
(f) $R = 30R_\star$ and $\alpha = 1$; for three inclination angles of the direction towards the 
observer in respect to the disk plane: $\theta = 15^\circ$ (solid curves), $45^\circ$ (dashed), 
and $90^\circ$ (dotted). 
The dependence of the $\gamma$-ray spectra on the distance from the star, $R = 2.5R_\star$ (solid), 
$5R_\star$ (dashed), $15R_\star$ (dotted) and $30R_\star$ (dot-dashed), for two
selected inclination angles of the observer, $\theta = 15^\circ$ and $90^\circ$, is shown in panels (c) and (d) for $\alpha = 2$ and (g) and (h) for $\alpha = 1$, respectively. 
The synchrotron energy losses of electrons, during their advection process along
the equatorial wind region, are taken into account. The magnetic field strength in the equatorial 
wind is derived assuming its global dipole structure described by $B = 0.5 \times B_{\star}
(R_\star/R)^3$.  The spectrum of electrons is normalized to $10\%$ of the part of the wind power, 
$L_{\rm w} = 1.56\times 10^{36}$ erg~s$^{-1}$, which arrive to the equatorial plane at 
the ring with the inner radius $R$ and thickness $0.2R_\star$.}
\label{fig2}
\end{figure*}

At first, we calculate the $\gamma$-ray spectra produced by electrons which are injected at 
the ring defined by the specific range of distances from the star. According to the considered 
scenario, 
electrons are accelerated in the reconnection regions to larger energies closer to the star 
due to the strong dependence of the dipole magnetic field on the distance from the star. 
However, $\gamma$-rays, produced by 
them in the IC process closer to the star, are also more 
efficiently absorbed in the stellar radiation. Therefore, the $\gamma$-ray spectra show 
complicated behaviour with the distance from the star.
On Figs.~2a,b and 2e,f, we investigate dependence of the IC $\gamma$-ray spectra on the 
location of the external observer in respect to the plane of the equatorial wind for two 
models of the spectrum of electrons, described by the spectral index 
equal to $\alpha = 2$ (2a,b) and $\alpha = 1$ (2e,f). The $\gamma$-ray spectra, produced closer 
to the star (the case for $R = 2.5R_\star$), show strong dependence on the inclination 
angle due to strongly dependent conditions for their absorption in the stellar radiation. Only 
a part of the TeV $\gamma$-rays is able to escape from the stellar radiation field. 
The most favourite conditions for the escape of $\gamma$-ray photons are from the part of 
the equatorial disk which lays in front of the star in respect to 
the external observer (the interaction angle between the $\gamma$-ray photon and the stellar 
photon is the largest). On the other hand, $\gamma$-ray spectra, produced at large distances 
from the star, extend only to sub-TeV energies. Their level is within 
a factor of two for the range of angles between $\theta = 15^\circ - 90^\circ$ (see Fig.~2bf 
for $R = 30 R_\star$). The weakest spectra are observed for small inclination angles of the observer.
In fact, weak dependence of the $\gamma$-ray spectra on the observation angle, $\theta$, is not so 
surprising. They are produced in specific parts of the ring as a result of the IC scattering of 
stellar radiation arriving at quite large range of angles. 
Therefore, $\gamma$-ray emission from the whole ring is averaged over large range of collision 
angles. In Figs.~2, we also show the dependence of the $\gamma$-ray spectra on the distance from 
the star for two fixed inclination angles, $\theta = 15^\circ$ (Figs 2c,g) 
and $90^\circ$ (Figs. 2d,h). We show that $\gamma$-ray spectra are produced
with the largest energies, extending to the TeV range, at the intermediate distances from the star 
(for example see the case for $R\sim 5R_\star$). Instead, electrons are accelerated to the largest 
energies at the inner radius of the equatorial disk. This effect is due to efficient absorption 
of $\gamma$-rays in the stellar radiation,
with energies close to $\sim 100$ GeV Also, the dominant synchrotron energy losses of electrons 
(in comparison to the IC losses) in the inner part of the equatorial disk plays an essential role
in drawing energy from electrons at the inner radius of the equatorial disk.

\begin{table}
  \caption{Parameters of the models for electron acceleration}
  \begin{tabular}{llllll} 
\hline 
\hline 
 model:  & $\alpha$  & $\xi$ &  $E_{\rm max}(R_{\rm in}=2.4R_\star)$  & $E_{\rm max}(R_{\rm out}=30R_\star)$   \\
\hline
I        &   2         & 1     &  40 TeV  &  260 GeV  \\
\hline
II       &   1         & 1     &  40 TeV  &  260 GeV  \\
\hline
III     &    1         &  0.1  &   4 TeV  &  26 GeV \\
\hline 
\hline 
\end{tabular}
  \label{tab2}
\end{table}

\begin{figure*}
\vskip 9.truecm
\includegraphics{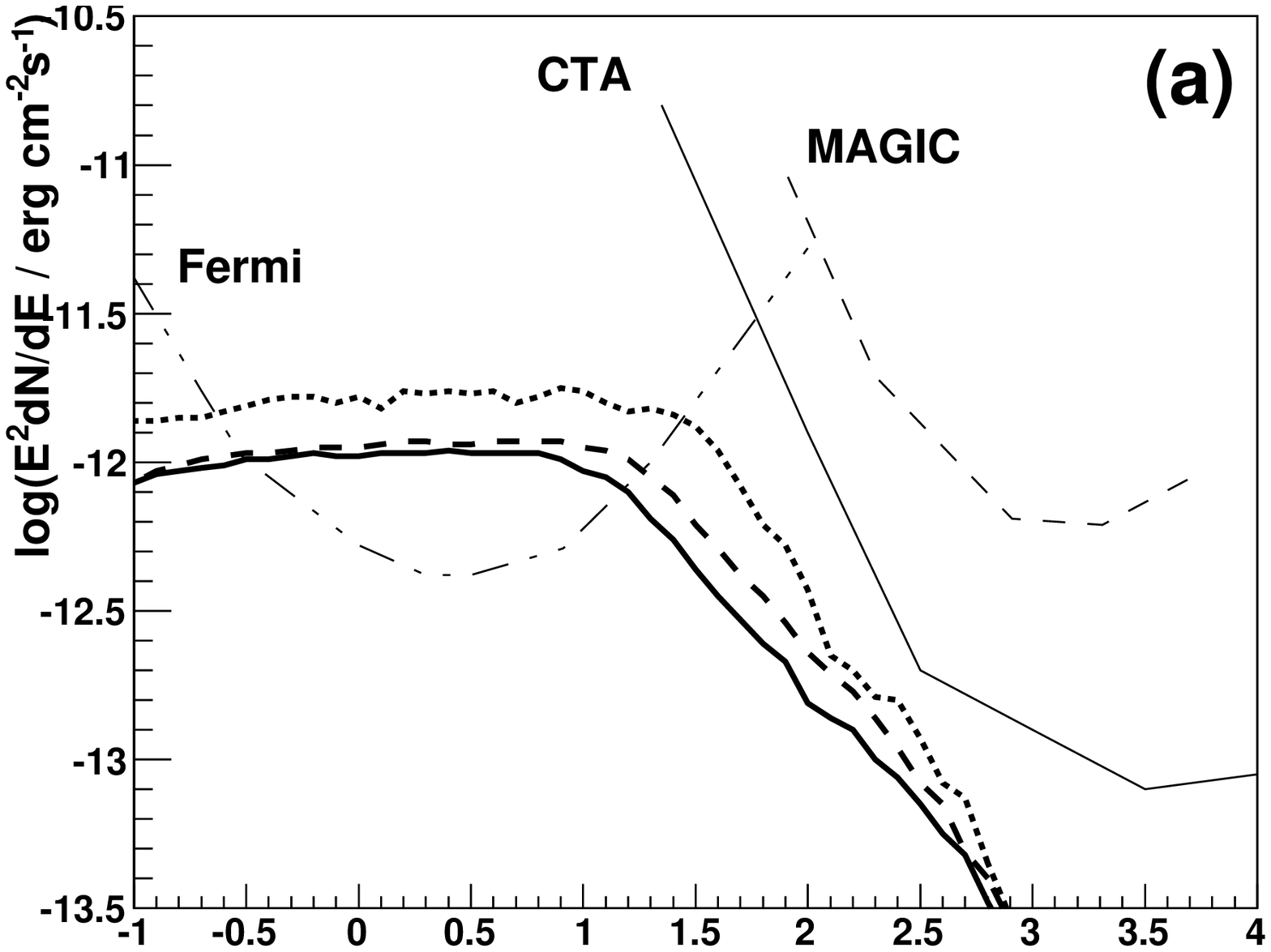}
\includegraphics{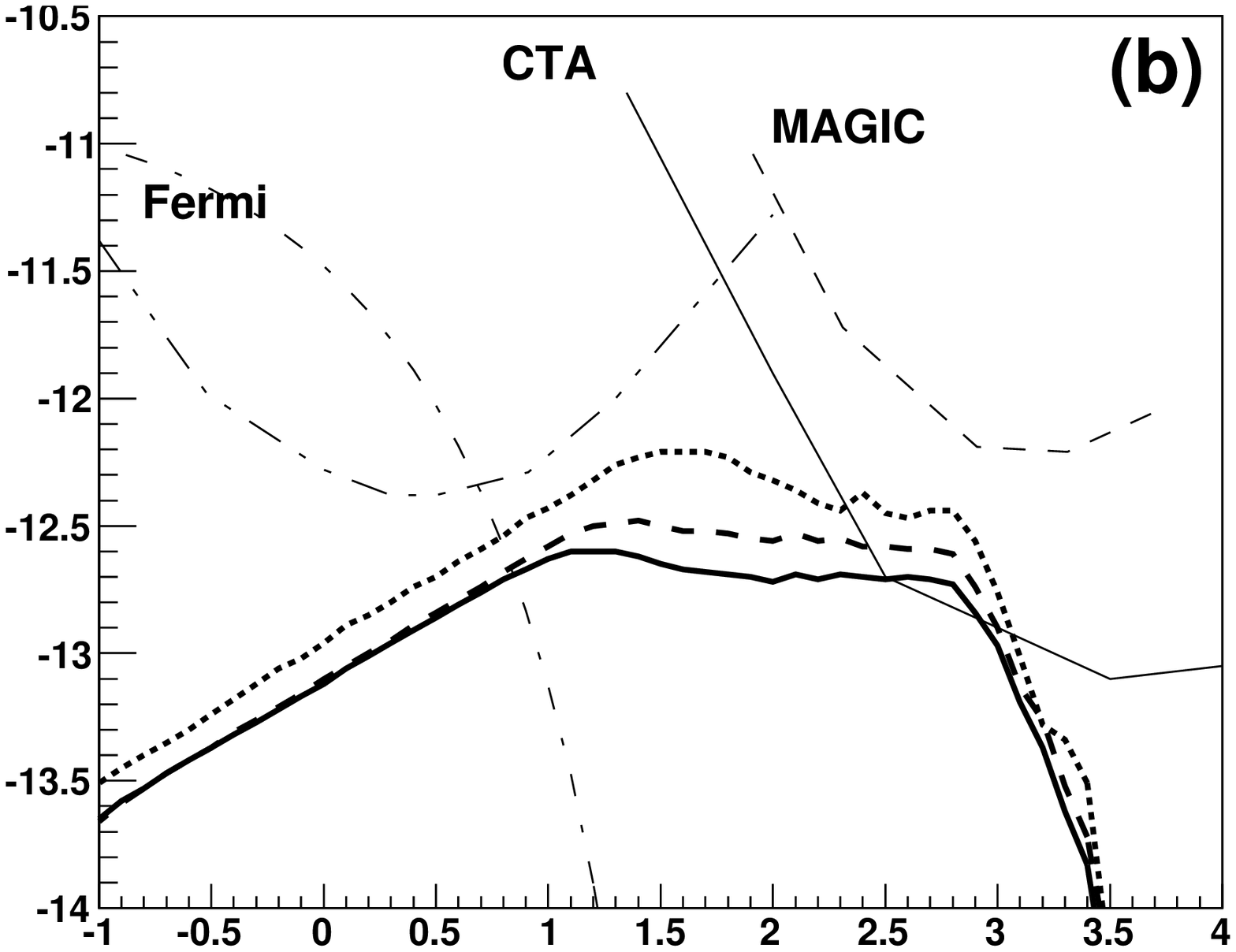}
\includegraphics{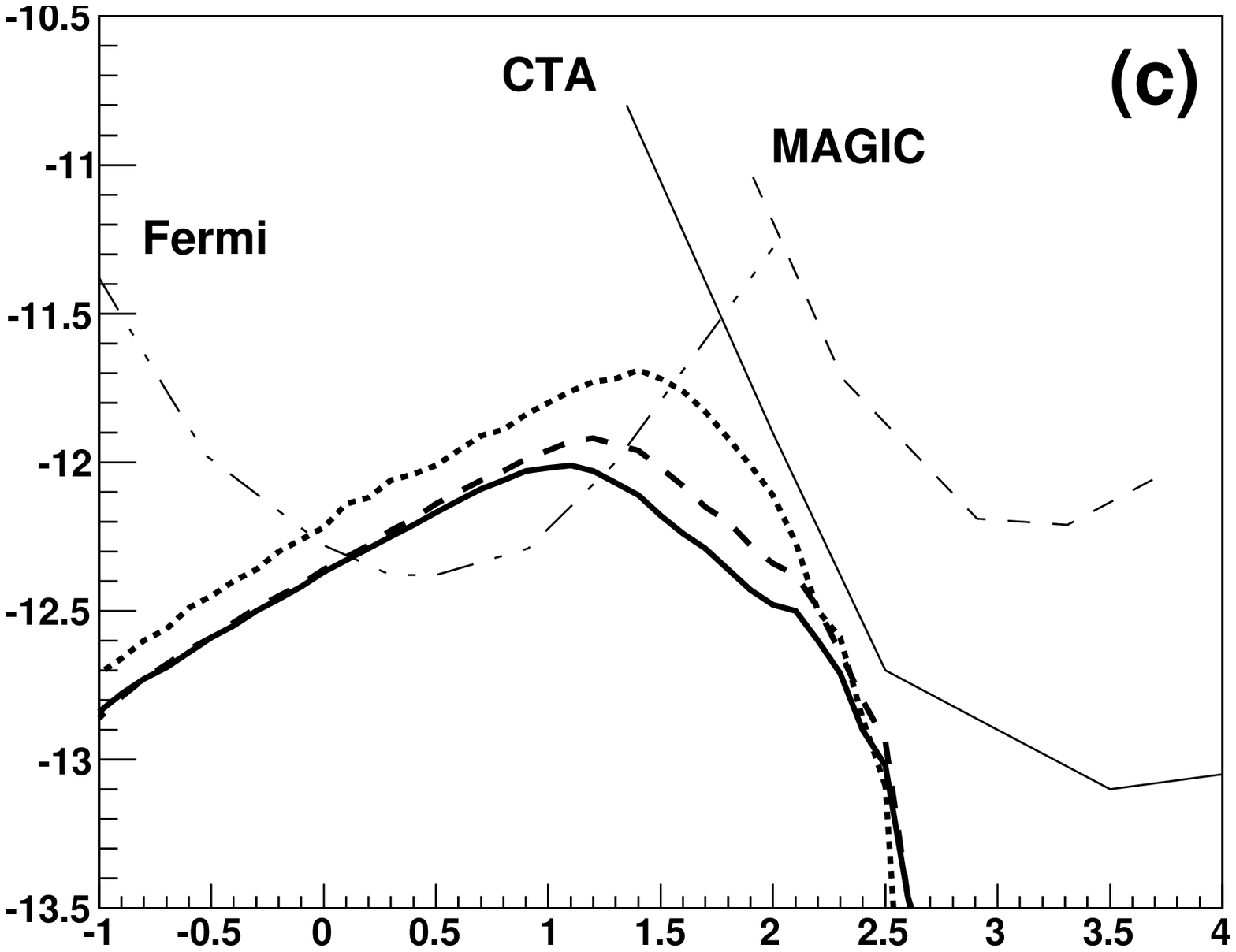}
\includegraphics{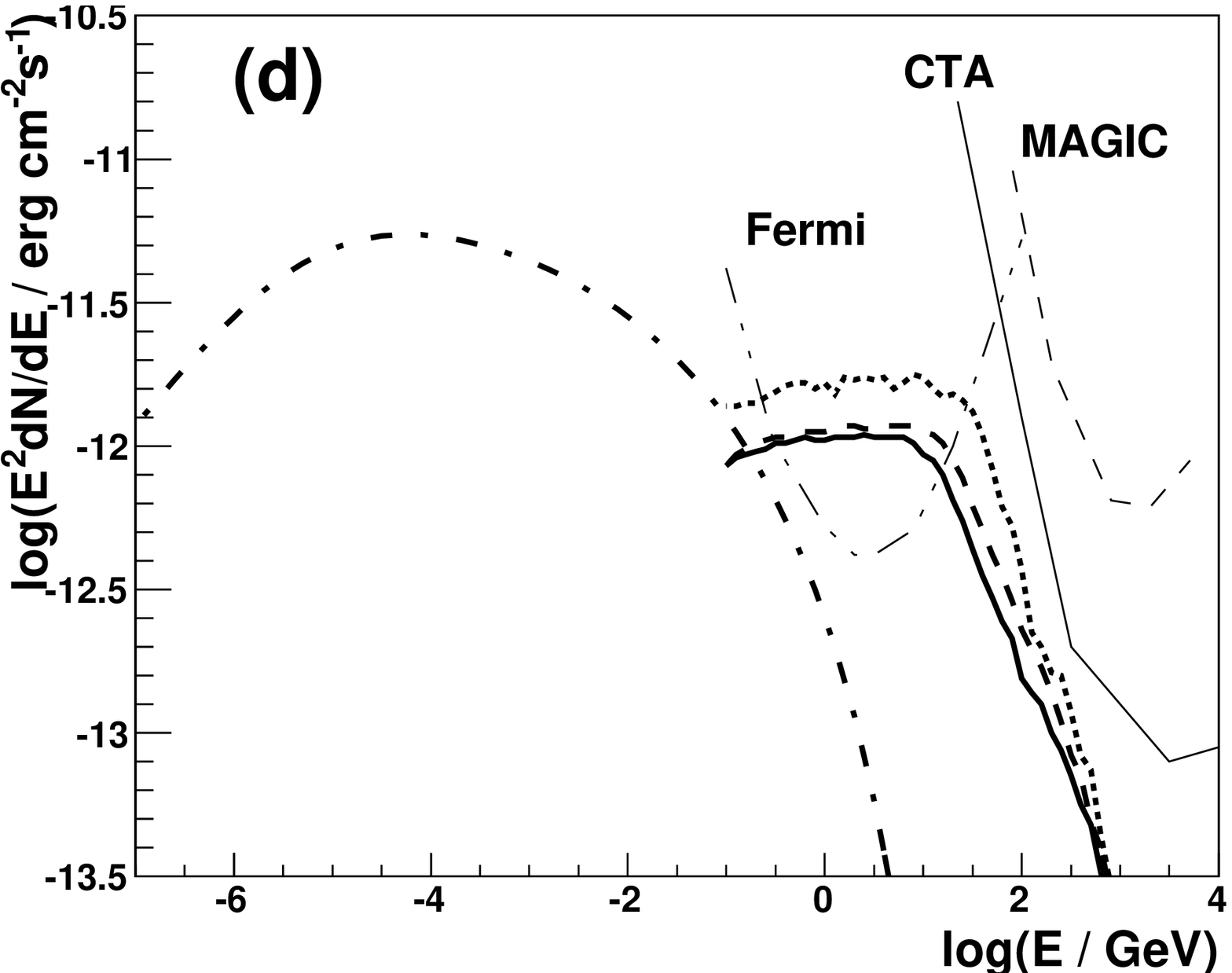}
\includegraphics{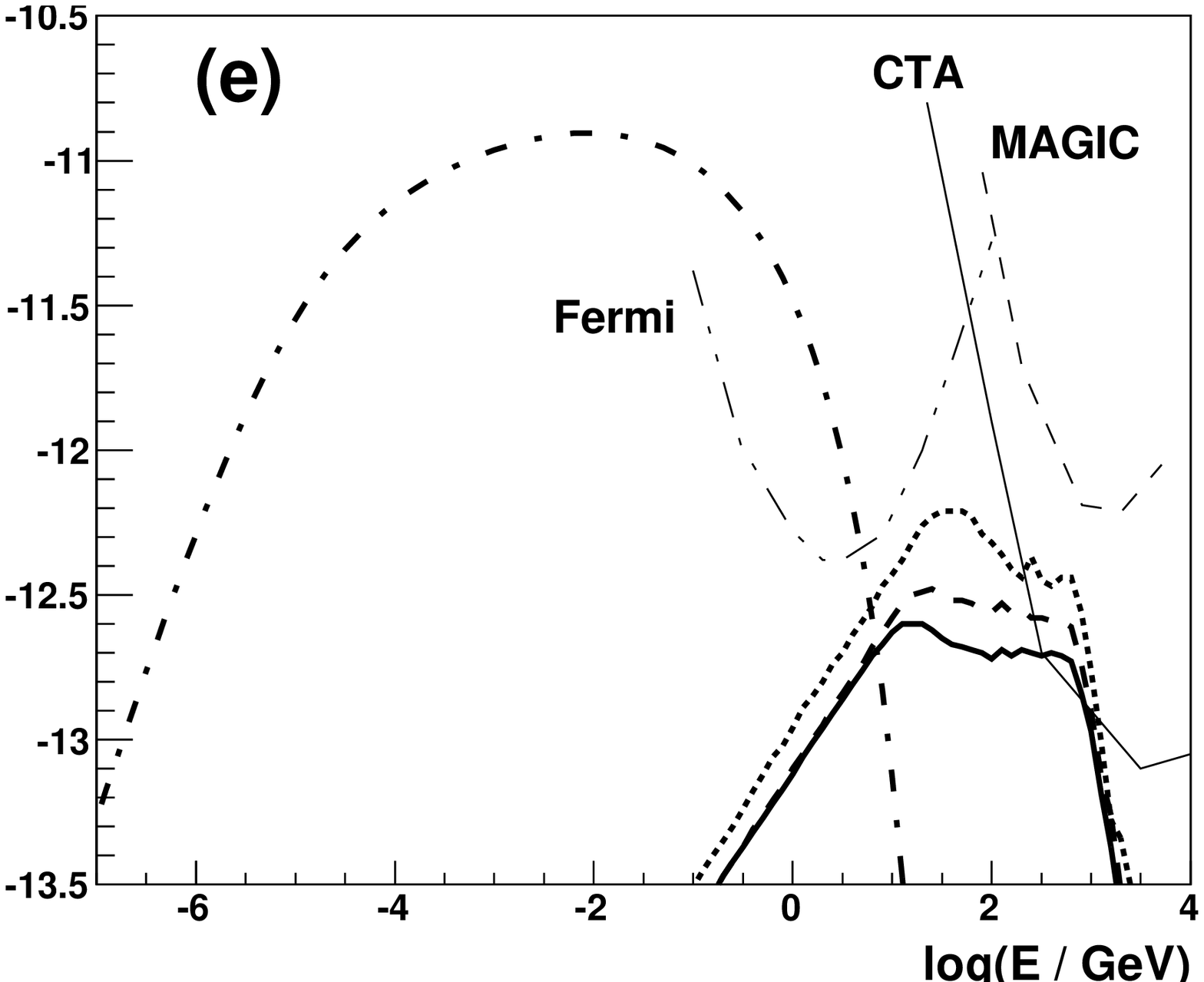}
\includegraphics{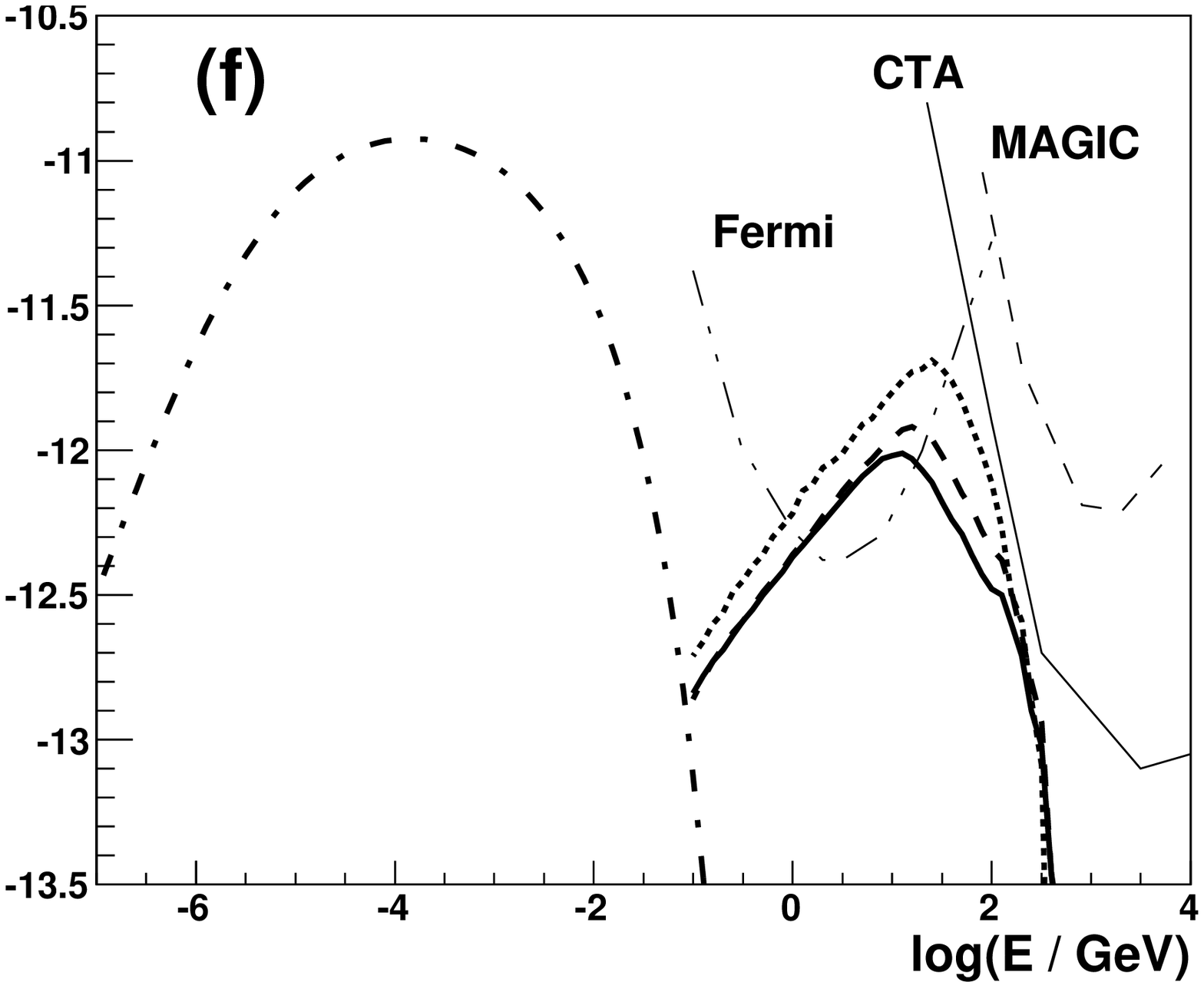}
\caption{Gamma-ray spectra produced by relativistic electrons in the equatorial wind region 
between the range of distances from the star $R_{\rm min} = 2.5 R_\star$ and 
$R_{\rm max} = 30 R_\star$. Electrons IC up-scatter thermal radiation from the star with 
the parameters of HD 37022. Electrons are injected, from the reconnection 
regions in the equatorial wind, with the power law spectrum defined by the spectral index 
$\alpha$ and maximum energies $E_{\rm max}$. 
The synchrotron energy losses of electrons, during their advection process along 
the equatorial wind, are included. Three models for the electron acceleration and considered. 
Model I: the spectral index is $\alpha = 2$ and the length of the reconnection region is 
$\xi = 1$, see figure (a) for the synchrotron and the IC spectra and (d) for details of IC spectra. 
Model II: $\alpha = 1$ and $\xi = 1$ (figures (b) and (e), respectively). Model III: 
$\alpha = 1$ and $\xi = 0.1$ (figures (b) and (e) respectively).
Specific lines show the results for three inclination angles of the distant observer in respect to the plane of the equatorial disk: $\theta = 15^\circ$ (solid curve), 
$45^\circ$ (dashed), and $90^\circ$ (dotted).
It is assumed that the magnetic field at the equatorial wind region is defined by 
the dipole structure.  
The spectrum of electrons is normalized to $10\%$ of the wind power, 
$L_{\rm w} = 1.56\times 10^{36}$ erg~s$^{-1}$, which arrive to the equatorial plane between 
$R_{\rm min}$ and $R_{\rm max}$. The gamma-ray spectra are confronted with the sensitivities of 
the Fermi-LAT (10 yrs, see dot-dot-dashed curve, Funk et al.~2013),  
the MAGIC array (50 hrs, thin dashed curve, Aleksi\'c et al.~2012) 
and the planned CTA array (50hrs, thin solid curve, Maier et al.~2017).}
\label{fig3}
\end{figure*}

In the next step, we calculate the $\gamma$-ray spectra produced by electrons in the IC process 
from the whole equatorial wind region. As an example, we select the disk with the inner radius 
corresponding to the Alfven radius in the magnetosphere of the star HD 37022 (i.e. $2.4R_\star$)
and the outer radius fixed 
on $30r_\star$. The outer radius is chosen in order to be consistent with the distance at 
which the structure of the magnetic field around the star changes to toroidal. At this distance, 
acceleration of electrons in the reconnection regions in the equatorial disk can still occur to 
energies of the order of  $\sim 100$ GeV (i.e. 
produced by them $\gamma$-rays can reach energies within sensitivities of the Cherenkov 
telescopes). We consider three models for the acceleration of electrons and for their subsequent 
energy losses during advection with the equatorial wind. In all models, electrons are 
injected from the reconnection regions with the power law spectrum with the exponential 
cut-off at energy $E_{\rm max}$. This maximum energy depends on the local magnetic 
field strength at the equatorial disk (i.e. depends on $R$) and on the length scale, $\xi$, 
of the reconnection region. Spectra of relativistic electrons are normalized to $10\%$ of 
the wind power which propagates towards the equatorial disk. 
The parameters of specific models are reported in Tab.~2.

\begin{figure*}
\vskip 9.truecm
\includegraphics{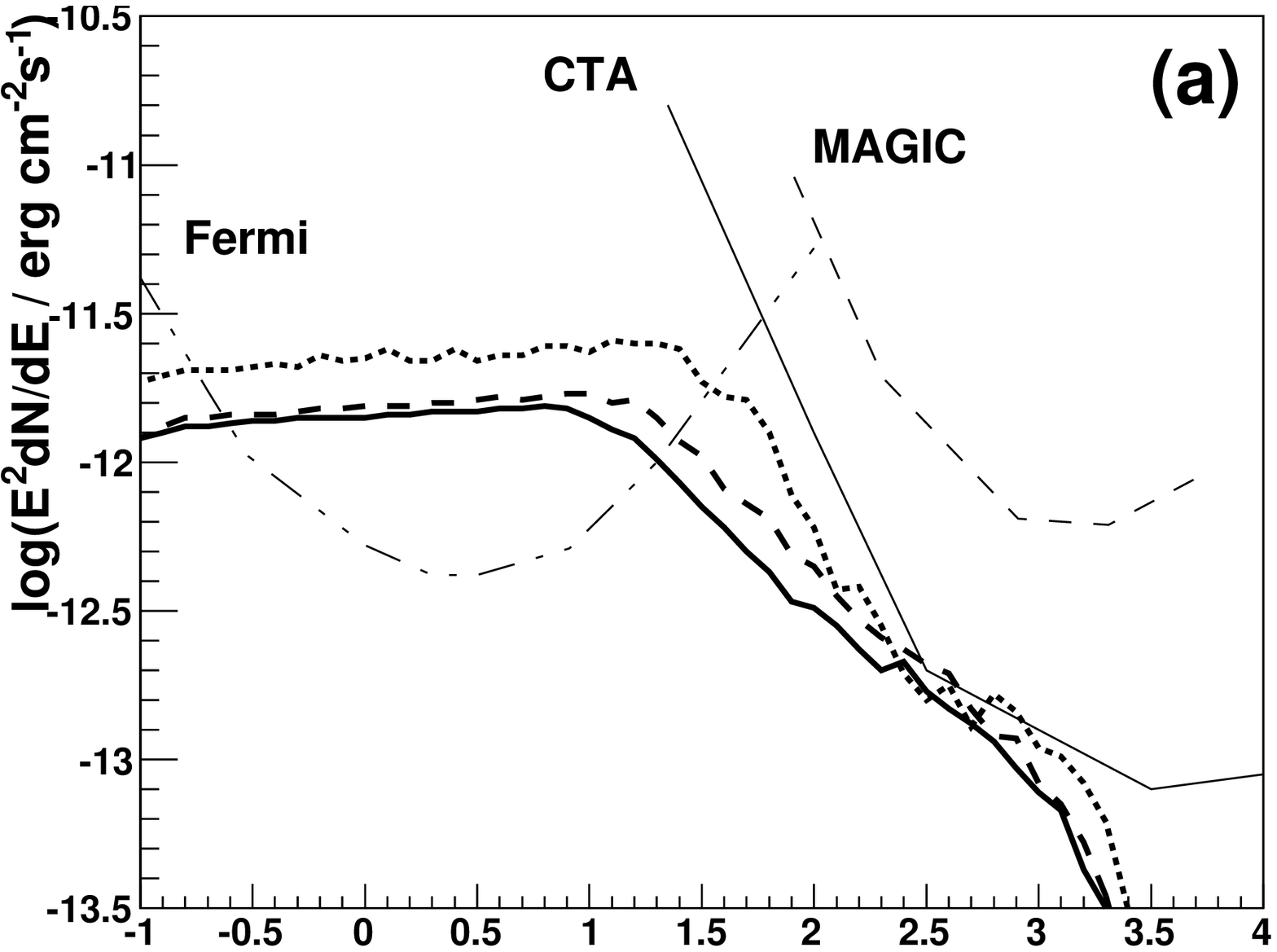}
\includegraphics{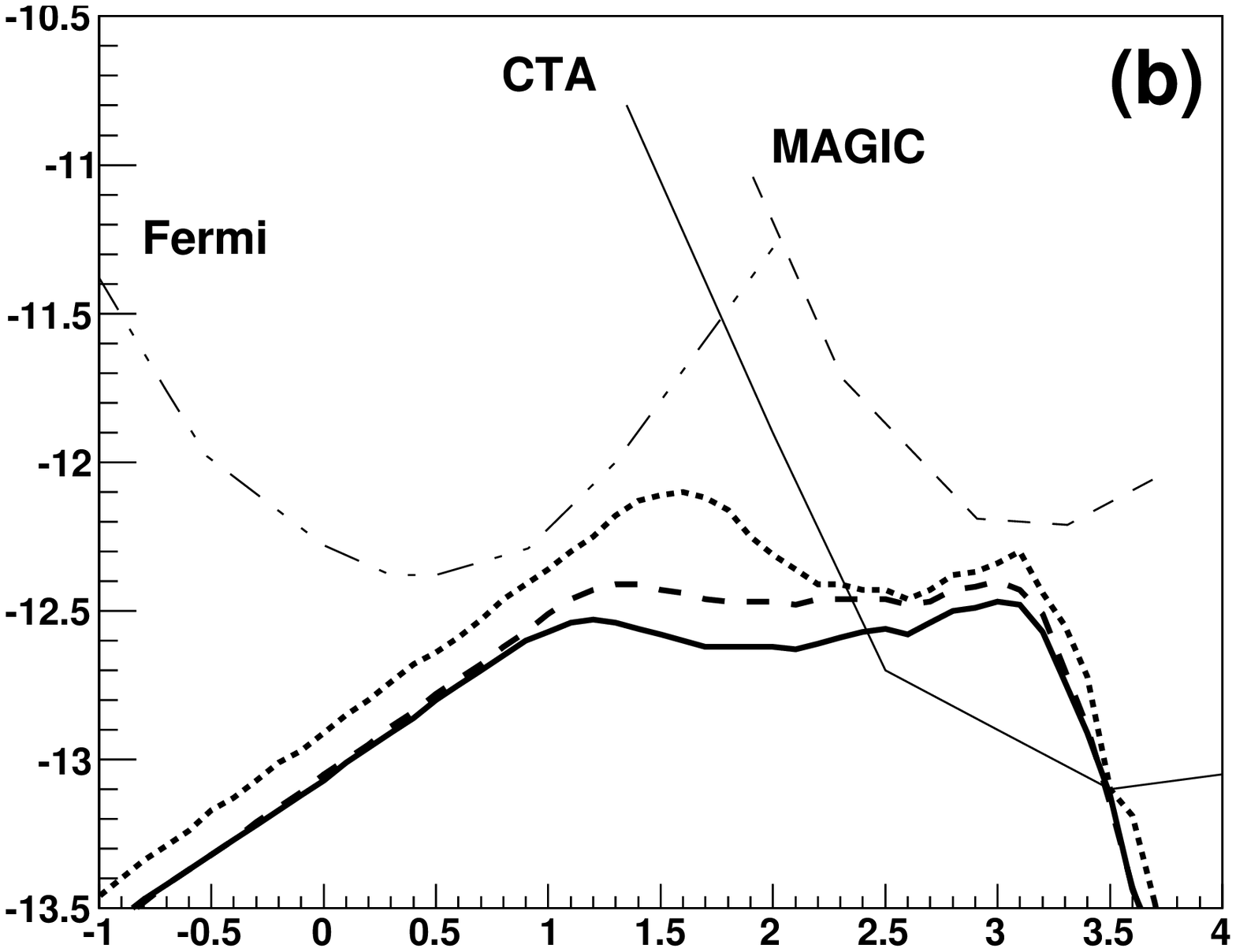}
\includegraphics{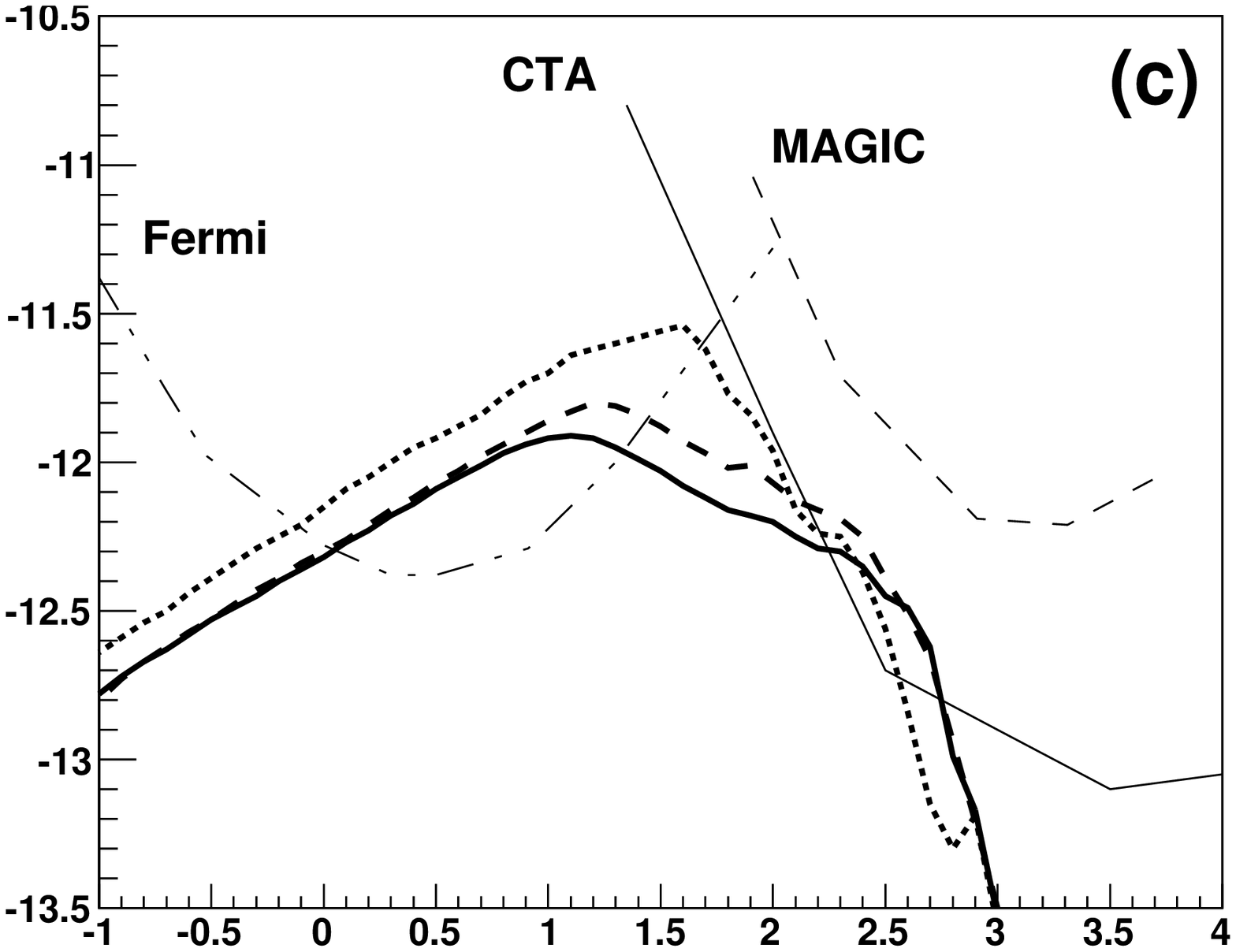}
\includegraphics{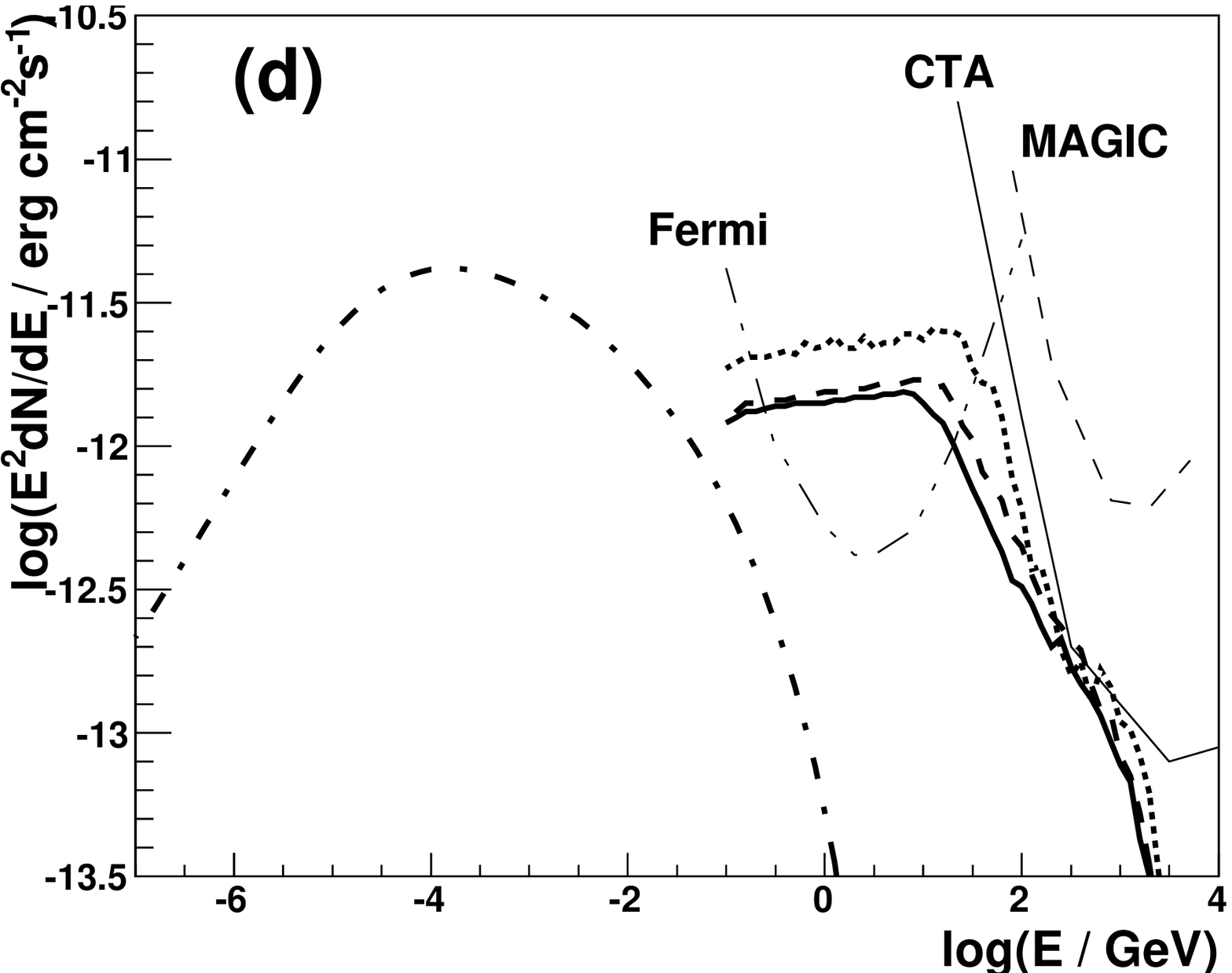}
\includegraphics{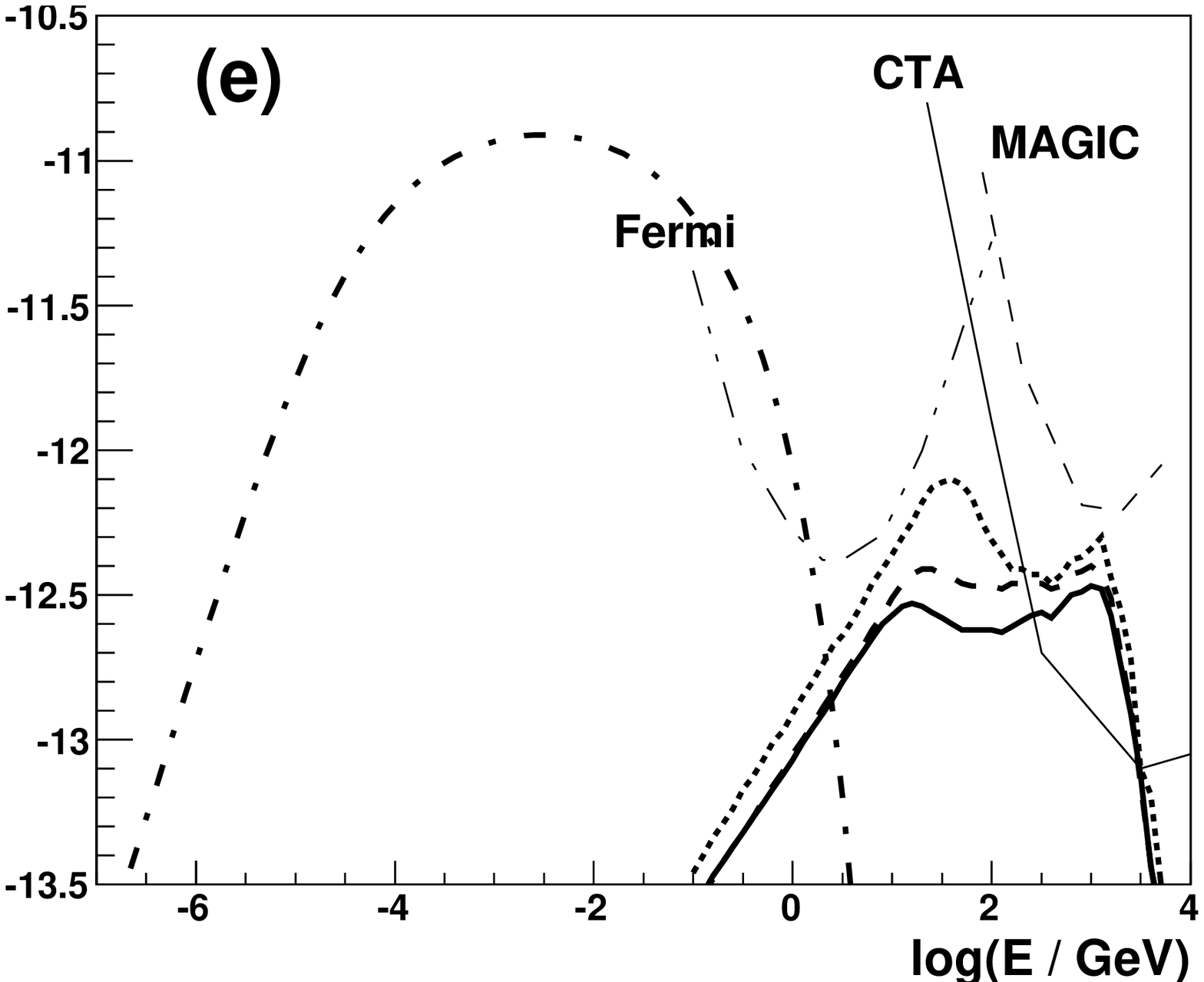}
\includegraphics{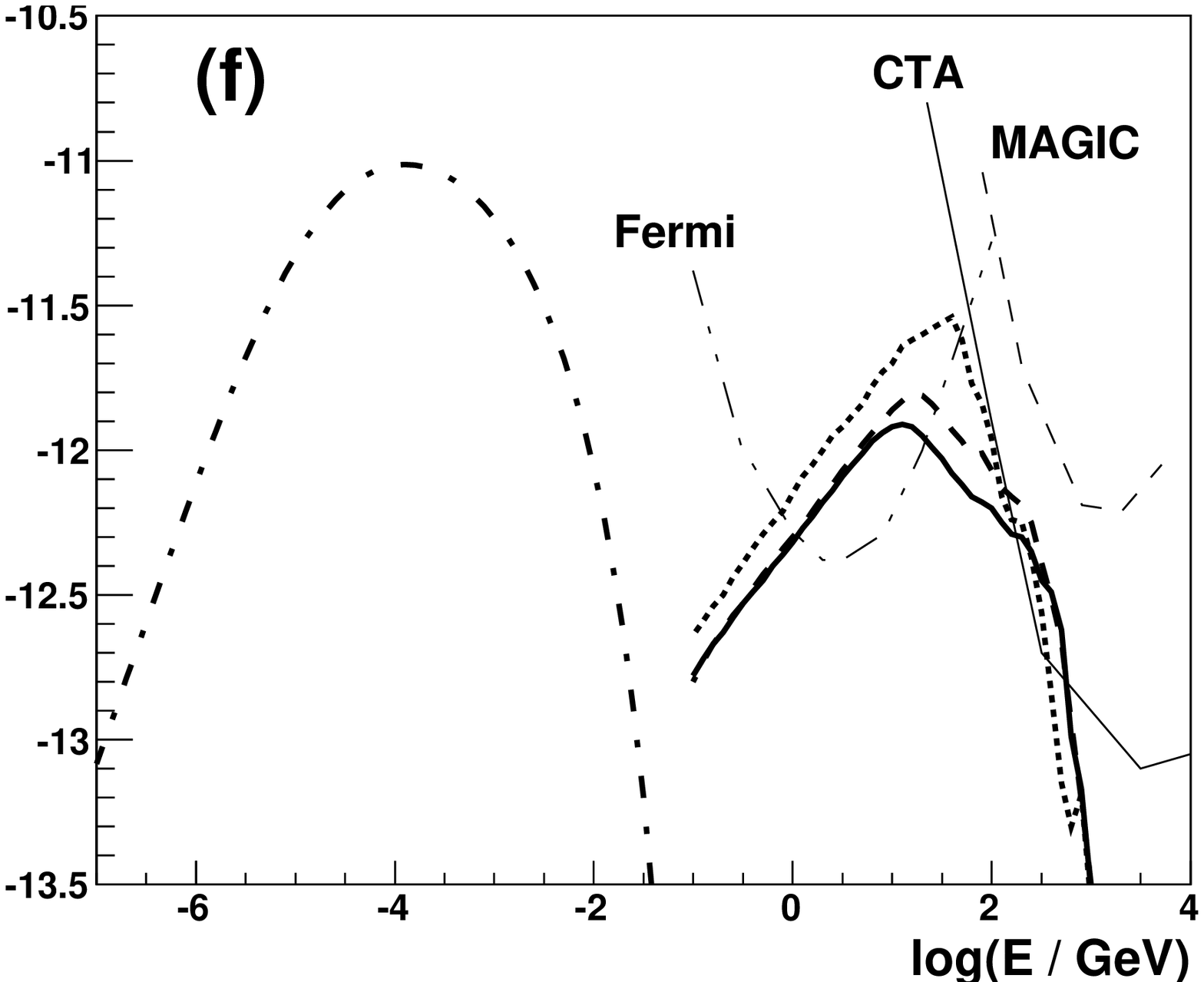}
\caption{As in Fig.~3 but for the reduction factor of the magnetic field strength in 
the equatorial wind equal to $\mu = 0.3$.}
\label{fig4} 
\end{figure*}
\begin{figure*}
\vskip 9.truecm
\includegraphics{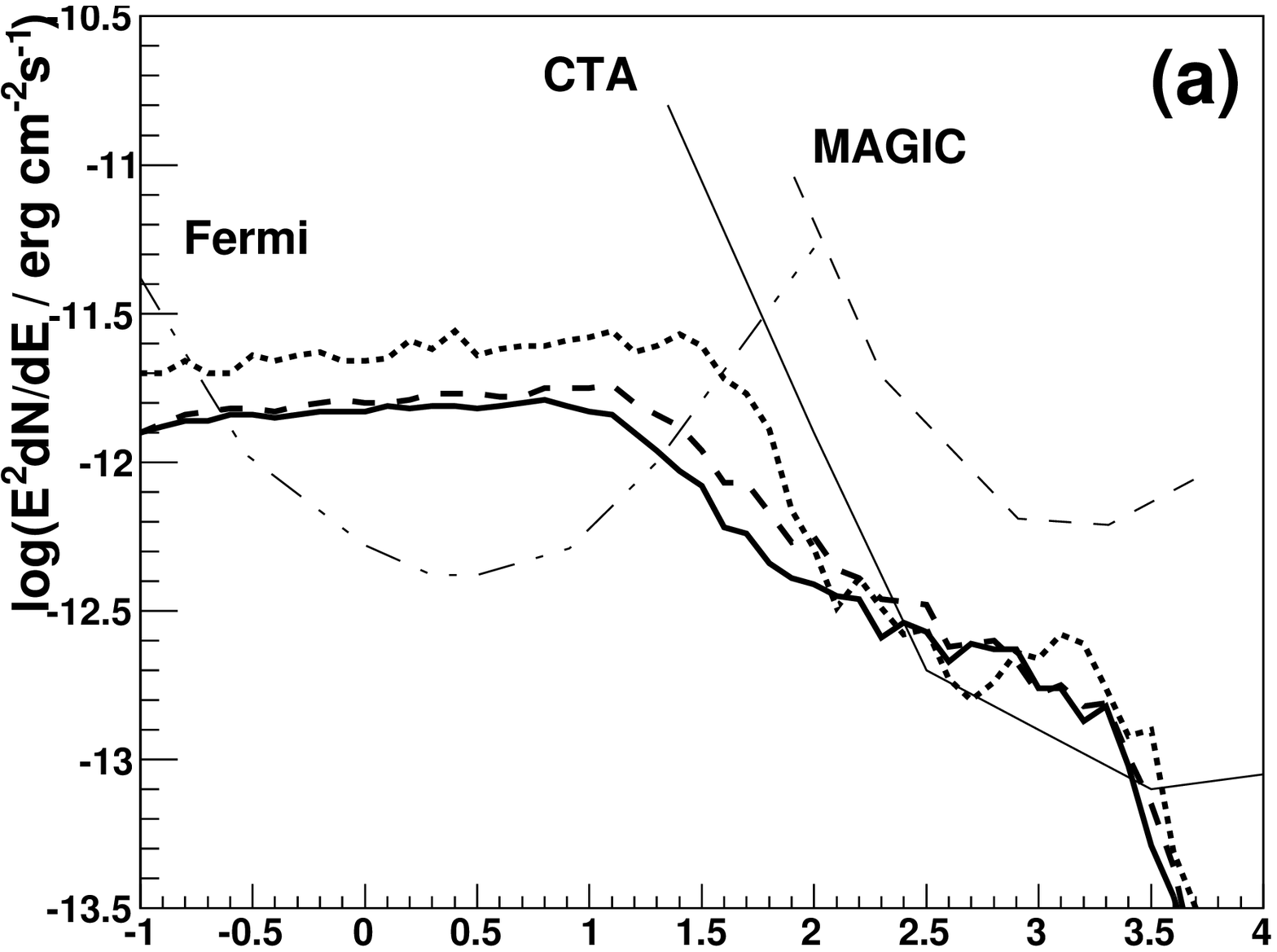}
\includegraphics{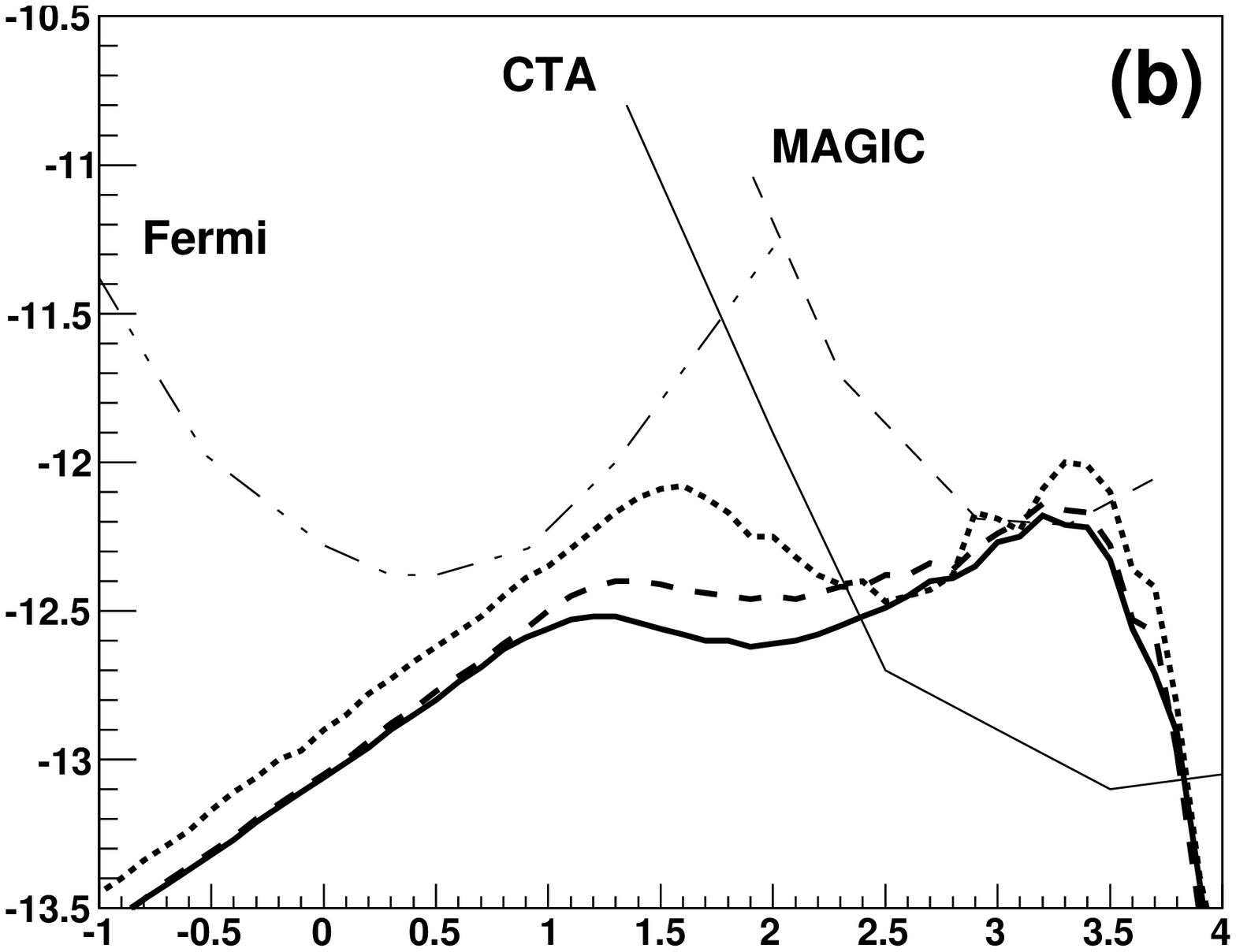}
\includegraphics{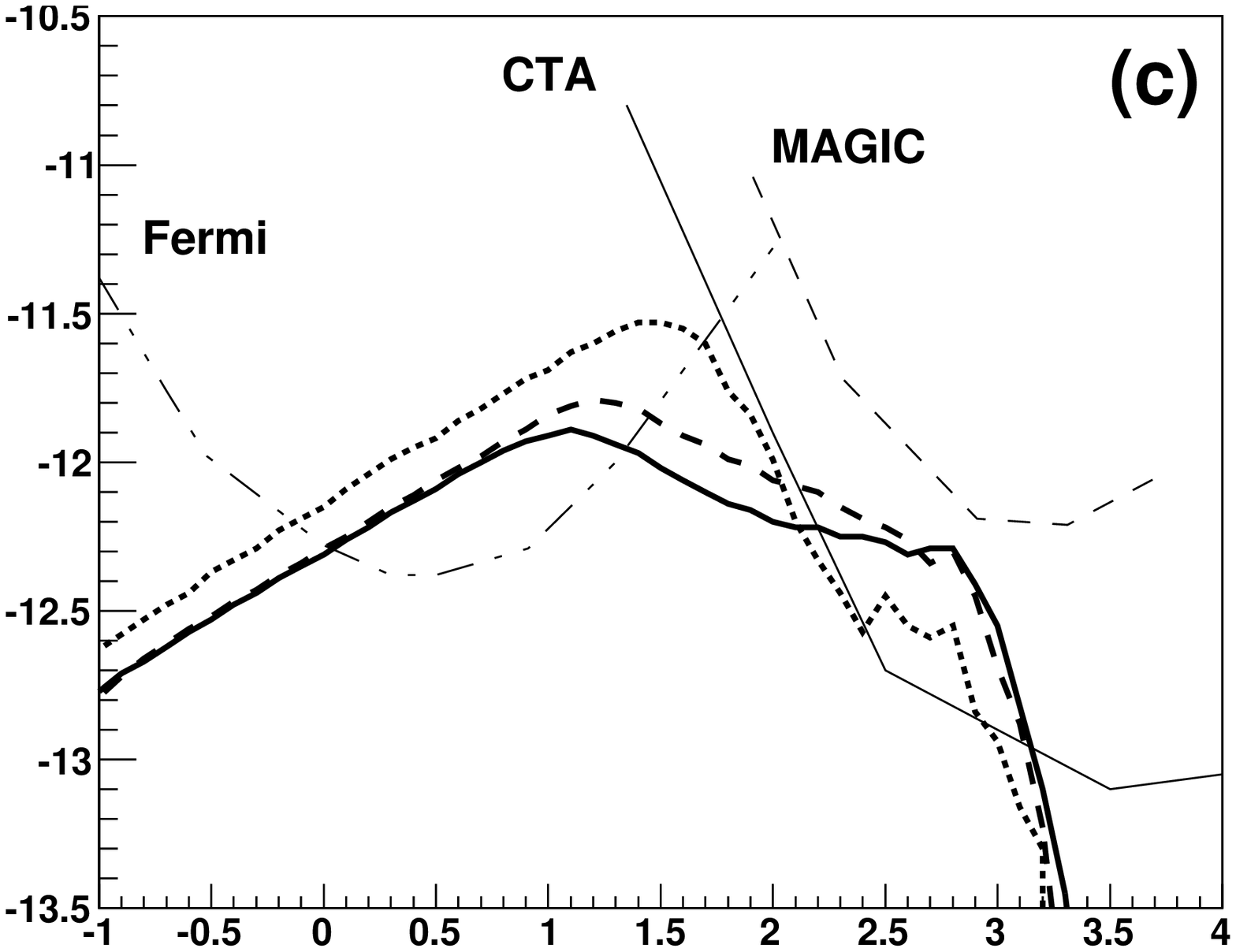}
\includegraphics{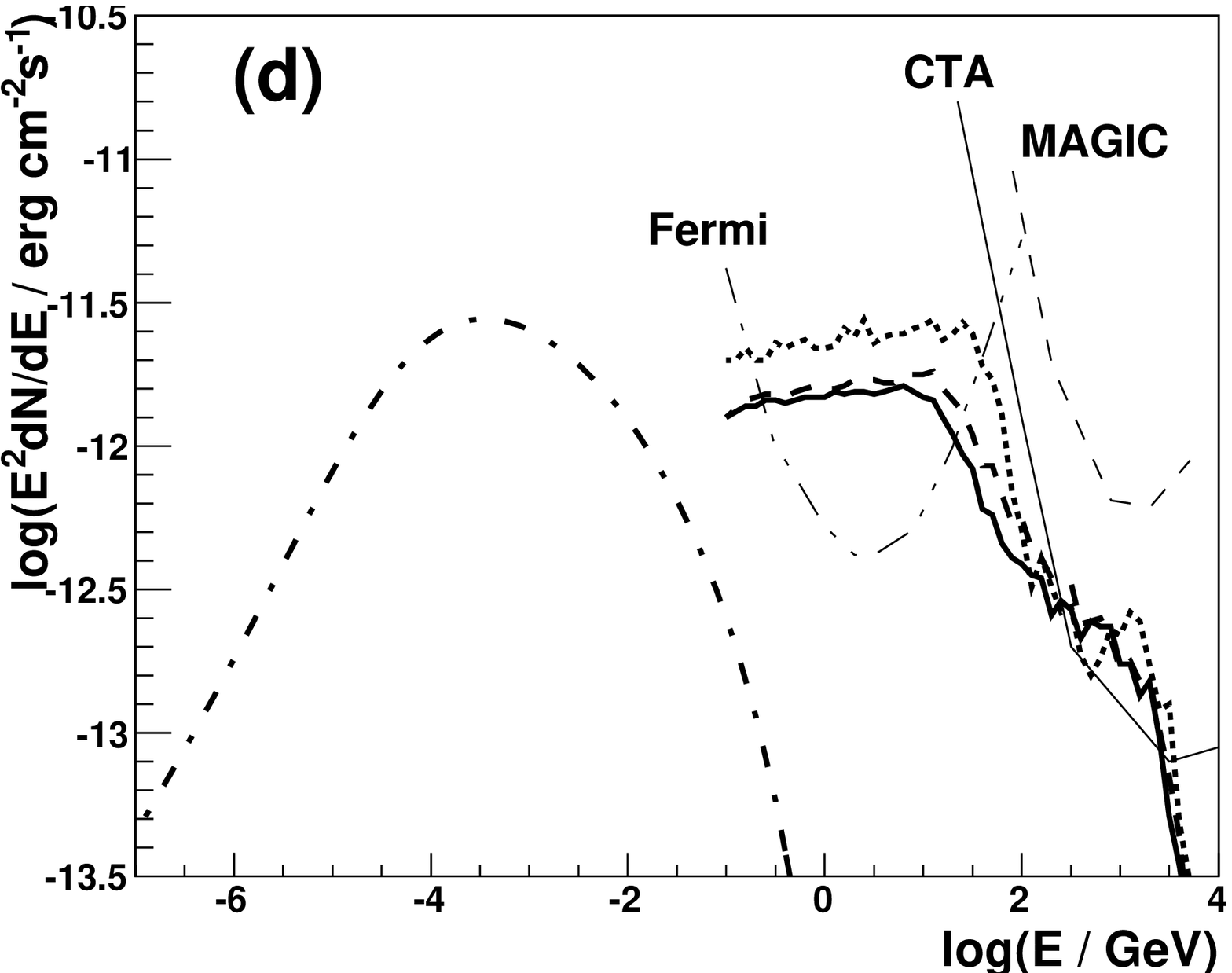}
\includegraphics{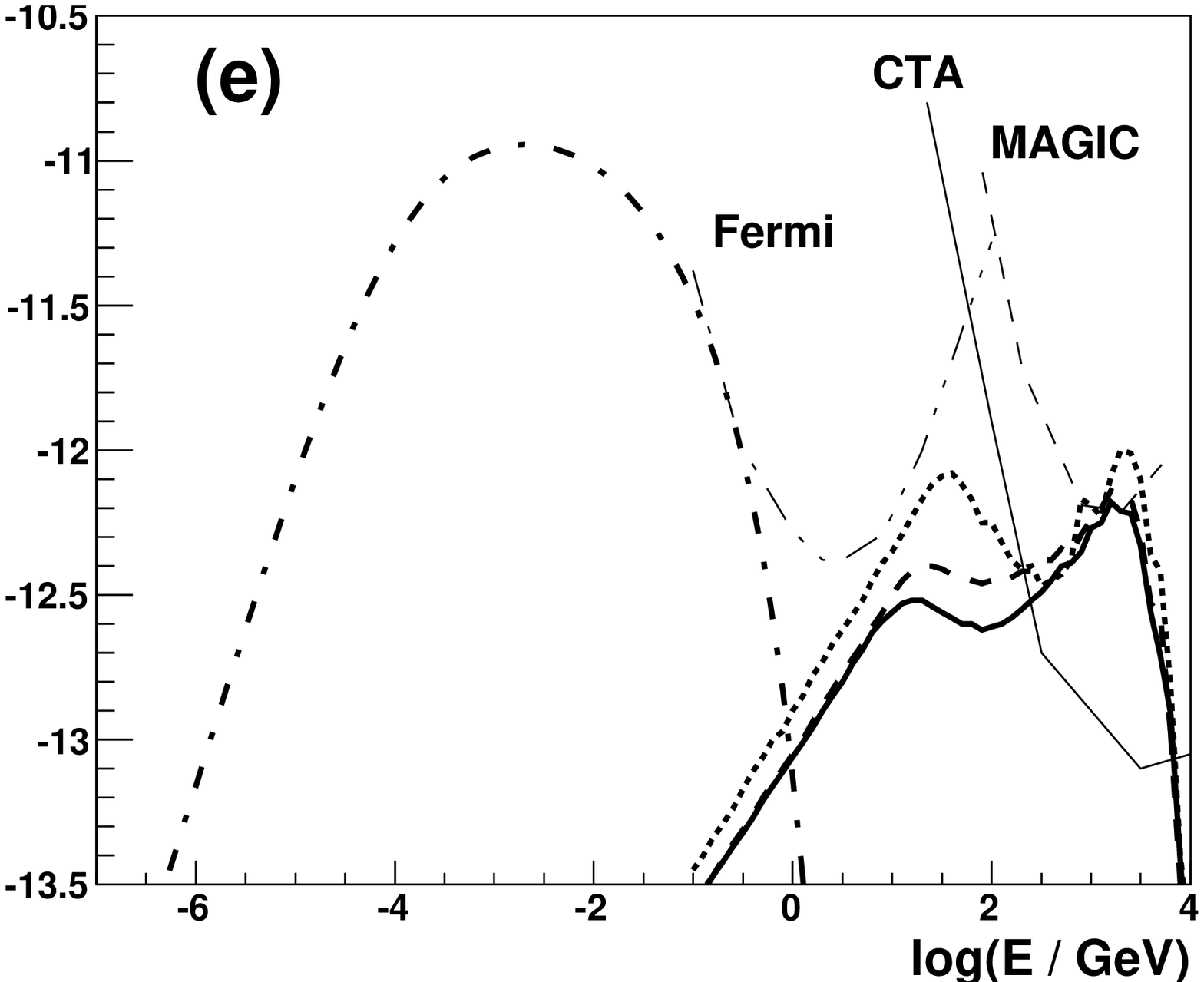}
\includegraphics{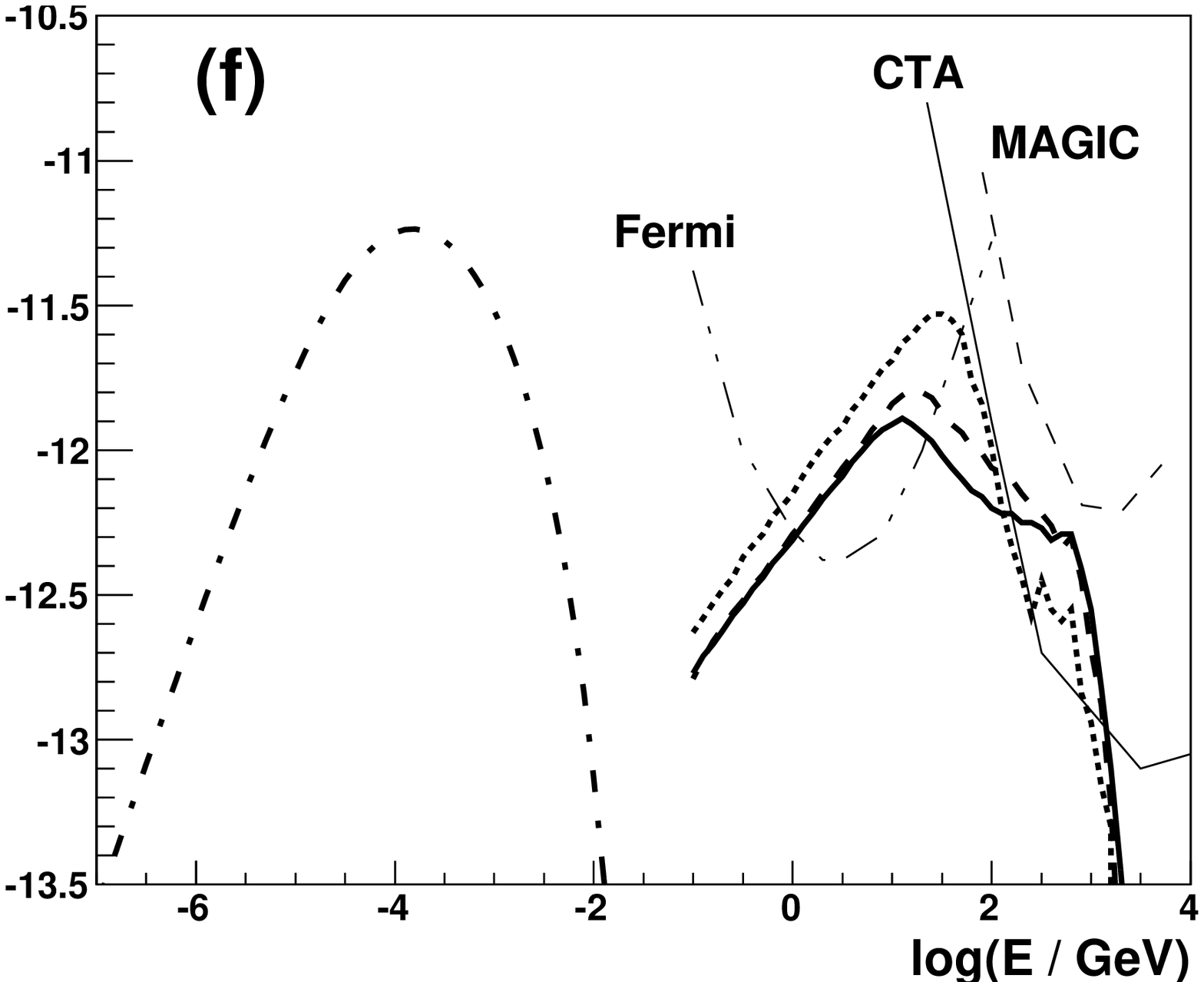}
\caption{As in Fig.~3 but for the reduction factor of the magnetic field strength in the 
equatorial wind region equal to $\mu = 0.1$.}
\label{fig5}
\end{figure*}

We perform calculations of the $\gamma$-ray and synchrotron spectra, 
in terms of the considered model for those three sets of example parameters. The spectra are 
shown in Fig.~3 for three inclination angles of the distant observer in respect to 
the plane of the equatorial disk: $\theta = 15^\circ$ (solid curve), $45^\circ$ (dashed), 
and $90^\circ$ (dotted). The results are confronted with the sensitivities of the present 
(Fermi-LAT , MAGIC) and future (CTA) $\gamma$-ray telescopes. In the case of the model I 
(Fig.~ 3ad), the $\gamma$-ray spectra extend through the GeV energy range with the fast 
decline above $\sim 30$ GeV. Such emission can be detected by
the Fermi-LAT telescope. However, the sub-TeV emission is clearly below the sensitivity of 
the future CTA. On the other hand, in the model II (Fig.~3be), the GeV $\gamma$-ray emission 
is predicted 
to be below sensitivity of the Fermi-LAT, but it might be marginally detected at energies 
close to $\sim 1$ TeV by the CTA.
The injection of electrons with the hard spectrum, but lower acceleration efficiency 
(model III, Fig.~3cf), turns to the $\gamma$-ray emission which is expected to have features 
similar to those presented in model I. However, the GeV
spectrum is flatter, steepening more suddenly in the sub-TeV range.  

The synchrotron emission is expected to dominate over the IC $\gamma$-ray emission in 
all considered models.  Interestingly, in the model II, the tail 
of the synchrotron emission extends up to GeV energies. This part of the synchrotron spectrum dominates completely over a part of the spectrum produced in the IC process.

In fact, strong thermal X-ray emission is observed from HD 37022 (Gagne et al.~2005).
It has been predicted to originate in 
the inner (closed) part of the magnetosphere, in which the winds, launched from the regions of 
magnetic poles, should collide in the equatorial region. As a result of this collision, 
strong, soft X-ray emission 
is produced. This soft X-ray emission dominates the energy output of the star. 
Predicted by our model broad band non-thermal synchrotron component, extending from the soft X-rays up to $\gamma$-rays, is dominated by the thermal soft X-ray emission 
from the inner magnetosphere.  

In fact, magnetic field in the equatorial wind can be significantly reduced 
from the values delivered by the stellar dipole magnetic field to the reconnection regions, 
since a part of the magnetic energy has to be transferred to relativistic particles and 
thermal plasma. 
Therefore, we also consider the cases in which electrons, injected from the reconnection 
regions into the equatorial wind, produce synchrotron and IC radiation in the region of 
significantly reduced magnetic field (by a factor of $\mu$). We investigate the synchrotron and 
the IC spectral features assuming that the reduction factor is equal to $\mu = 0.3$ (Fig.~4) and 
$\mu = 0.1$ (Fig.~5), for all 
three models of the electron injection, as considered in the case without magnetic field reduction 
(Fig.~3 calculated for $\mu = 1$). If the effect of the magnetic field reduction is important, then 
the synchrotron emission is predicted to be on a much lower level. On the other hand, 
the IC $\gamma$-ray emission
appears stronger. For $\mu= 0.1$, the tail of synchrotron emission is not expected to be 
detected by the Fermi-LAT telescope even in terms of the model II. On the other hand, 
sub-TeV $\gamma$-ray emission reaches the level allowing its detection by the CTA in the
case of all considered models. In the case of the model II, the peak of the IC $\gamma$-ray 
emission is even within the sensitivity of the present MAGIC telescope around TeV energies.

\section{Discussion and Conclusion}

We investigate the high energy radiation processes of relativistic electrons injected from 
the reconnection regions in the colliding winds around massive, magnetized star 
applying, as an example, the parameters of the star HD 37022 (known also as $\theta^1$ Ori C). 
Following previous investigations (e.g. Usov \& Melrose 1992, Trigilio et al.~2004), 
we assume that reconnection regions appear 
in the equatorial wind formed as a result of collisions of winds launched from the two 
magnetic poles of the star. Relativistic electrons are captured in the turbulent equatorial wind. 
They are advected with the wind in the outward direction from the star, 
suffering energy losses on the synchrotron process and on the IC scattering of stellar 
radiation. We calculate the synchrotron and the IC $\gamma$-ray spectra produced by these 
electrons in the case of a few models for the electron acceleration and propagation in the wind. 
We show that for the specific conditions, this $\gamma$-ray emission can be detected either in 
the GeV energy range by the Fermi-LAT telescope or by the future Cherenkov Telescope Array 
at sub-TeV energies, provided that close to $10\%$ of the wind power is transferred 
to relativistic electrons. If the acceleration process of electrons in the reconnection 
regions is very efficient (flat spectrum with spectral index close to $\alpha = 1$ and 
maximum energy at TeV energy range), then very steep high energy tail of the synchrotron emission might be 
even detected at GeV energies by the Fermi-LAT. 

$\gamma$-ray emission has been discovered in the GeV (and also TeV) energies from the direction of 
a few types of stellar objects (see e.g. Abdollahi et al.~2020). Two massive binary systems, Eta Carinae (e.g. Tavani et al.~2009, Abdo et al.~2010) and 
$\gamma^2$ Velorum (Pshirkov~2016) have been discovered in the GeV energy range. Eta Carinae has been also observed recently in the TeV energies (Abdalla et al.~2020). Moreover, Fermi-LAT and Agile telescopes detected GeV emission from a few binary systems 
containing compact objects. Some of them have been already discovered in the TeV energies. Between them, the best investigated are LSI 61+303 (Albert et al.~2006, Abdo et al.~2009a) and LS5039 (Aharonian et al.~2005a, Abdo et al.~2009b). 
In the case of another binary system, PSR1259-63/SS2883 (e.g. Tam et al.~2011, Aharonian et al.~2005b), the compact object is well documented as a rotation powered neutron star. 
To my knowledge any single, isolated star is observed in $\gamma$-rays up to now.

However, in the case of binary systems,  
proposed here model for the  high energy $\gamma$-ray emission can provide additional component to the 
high energy emission expected from a large scale shock formed as a result of collisions of 
stellar winds within the binary system (see e.g. Eichler \& Usov~1993, Benaglia \& Romero~2003, 
Bednarek~2005, Reimer et al.~2006, Pittard \& Dougherty~2006, Farnier et al.~2010, 
Bednarek \& Pabich~2011). In fact, both processes of electron acceleration 
(i.e. shock acceleration on the large scale double shock structure within the binary and
the reconnection in the wind of the magnetized star) can contribute to the emission from specific 
stellar binary systems and also binary systems of massive stars with compact objects (neutron stars, 
black holes). As argued in the above mentioned works, also hadrons can contribute to the $\gamma$-rays 
in the interaction with the matter 
of dense winds or in stellar atmospheres. Therefore, it is not surprising that 
the high energy emission from massive binary systems can show very complicated features. 
In fact, a few  emission components are expected to appear in different phases of the binary 
system, since all those 
processes depend significantly on their parameters and viewing geometry.      

In our calculations, we assumed steady conditions in the equatorial wind. However, the stellar winds 
are often non-stationary and time dependent. Moreover, the dipole magnetic field can vary on 
a longer time scale. In such a case, the conditions in the wind around the 
specific star can change drastically. This should have serious effect on the stability of 
predicted 
here features of the $\gamma$-ray emission. For more powerful winds, the inner radius of 
the equatorial disk is expected to shift towards the star. More power is expected to be 
transferred to relativistic electrons. However, electrons will lose energy more efficiently on 
the synchrotron process in stronger magnetic field. So then, produced IC $\gamma$-ray spectra are 
expected on a higher level but they are limited to lower energies, 
without significant sub-TeV component. 
If the dipole magnetic field of the star becomes weaker from some reasons, then the inner radius 
of the equatorial disk should also  move closer to the star. In such a case, synchrotron energy 
losses of electrons are expected on a similar level but more energy is transferred from the wind 
to electrons. Then, the 
$\gamma$-ray spectrum is expected to strengthen and shifts to higher energies. This should allow  
easier detection of the sub-TeV $\gamma$-ray emission by the Cherenkov telescopes.

Another complication is due to the inclination of the magnetic dipole axis to the rotational 
axis of the star. In such a case, the viewing angle of the equatorial disk by a distant observer 
changes regularly with the rotation period of the star (which is equal in the case of HD 37022 
to 15.4 days, Stahl et al.~1993). In the case od HD 37022, the magnetic dipole axis is inclined at  the angle 
$42^\circ$ and the observer is located at the angle $45^\circ$ in respect to the rotational 
axis of the star (Gagne et al.~2005). Therefore, the viewing angle of the equatorial disk by 
the observer
changes from $3^\circ$ (nearly pole on) up to $87^\circ$ (nearly equatorial disk on) during the 
rotational period of the star. For such parameters, we predict the periodic fluctuation of 
the level of $\gamma$-ray emission (mainly at a few tens of GeV) with the rotational period of 
the star by a factor close to $2-3$ (see spectra on Figs.~3-5).   

In the specific case of the example star, $\theta^1$ Ori C, the $\gamma$-ray emission discussed 
here is expected below the extended GeV $\gamma$-ray emission, recently observed 
from the Orion Molecular Complex by the Fermi-LAT and Agile satellites (Ackermann et al.~2012a,2012b, 
Marchili et al.~2018).  However, chances of detection 
of the TeV $\gamma$-ray emission by the Cherenkov telescopes, over the presumed extended 
background emission from the Orion Molecular Complex, are expected to be better due to superior 
angular resolution of the Cherenkov telescopes. In fact, recently predicted extended TeV $\gamma$-ray 
flux from the Orion Molecular Complex (Aharonian et al.~2020) is on the level of the TeV 
$\gamma$-ray flux expected in our calculations for HD 37022 ($\Theta^1$ Ori C).   
Recent upper limit on the TeV $\gamma$-ray emission from the Orion Molecular Complex, 
reported by the HAWC Observatory (Albert et al.~2021), is clearly above those predictions.

\section*{Acknowledgments}
I would like to thank the Referee for many useful comments.
This work is supported by the grant through the Polish National Research Centre 
No. 2019/33/B/ST9/01904.

\section*{Data Availability}
The simulated data underlying this article will be shared on
reasonable request to the corresponding author.


\label{lastpage}


\begin{thebibliography}{99}
\bibitem[Abdalla et al.(2020)]{abd20} Abdalla, H., Adam, R., Aharonian, F.  et al. 2020 A\&A 635, 167
\bibitem[Abdo et al.(2009a)]{abd09a} Abdo, A.A., Ackermann, M., Ajello, M. et al. 2009a ApJ 701, 123L
\bibitem[Abdo et al.(2009b)]{abd09b} Abdo, A.A., Ackermann, M., Ajello, M. et al. 2009b ApJ 706, L56
\bibitem[Abdo et al.(2010)]{abd10} Abdo, A.A., Ackermann, M., Ajello, M. et al. 2010 ApJ 723, 649
\bibitem[Abdo et al.(2011)]{abd11} Abdo, A.A., Ackermann, M., Ajello, M. et al. 2011 ApJ 734, 116
\bibitem[Abdollahi et al.(2020)]{abd202} Abdollahi, S., Acero, F., Ackermann, M. et al. 2020 ApJS 247, 33
\bibitem[Ackermann et al.(2011)]{ack11} Ackermann, M., Ajello, M., Allafort, A., et al. 2011 Science 334, 1103
\bibitem[Ackermann et al.(2012a)]{ack12a} Ackermann, M., Ajello, M., Allafort, A., et al. 2012a ApJ, 756, 4
\bibitem[Ackermann et al.(2012b)]{ack12b} Ackermann, M., Ajello, M., Allafort, A., et al. 2012b ApJ, 755, 22
\bibitem[Aharonian et al.(2005a)]{aha05a} Aharonian, F., Akhperjanian, A. G., Aye, K. -M. et al. 2005a Science 309 746
\bibitem[Aharonian et al.(2005b)]{aha05b} Aharonian, F., Akhperjanian, A. G., Aye, K. -M. et al. 2005b A\&A 442 1
\bibitem[Aharonian et al.(2020)]{aha20} Aharonian, F., Peron, G., Yang, R., Casanova, S., Zanin, R. 2020 PRD 101, 083018
\bibitem[Albert et al.(2006)]{alb06} Albert, A., Aliu, E., Anderhub, H. et al. 2006 Science 312 1771
\bibitem[Albert et al.(2021)]{alb21} Albert, A., Alfaro, R.; Alvarez, C. et al. 2021 ApJ 914, 106
\bibitem[Aleksi\'c et al(2012)]{ale12} Aleksi\'c, J., Alvarez, E. A.; Antonelli, L. A. et al. 2012, APh 35, 435
\bibitem[Babel \& Montmerle(1997)]{bm97} Babel, J., Montmerle, T. 1997 ApJ 485, L29 
\bibitem[Bednarek(1997)]{bed97} Bednarek, W. 1997 A\&A 322, 523 
\bibitem[Bednarek(2005)]{bed05} Bednarek, W. 2005 MNRAS 363, 46
\bibitem[Bednarek(2006)]{bed06} Bednarek, W. 2006 MNRAS 368, 579
\bibitem[Bednarek \& Pabich(2011)]{bp11} Bednarek, W., Pabich, J. 2011 A\&A 530, 49
\bibitem[Benaglia \& Romero(2003)]{br03} Benaglia, P. \& Romero, G.E. 2003 A\&A 399, 1121
\bibitem[Branduardi-Raymont(2007)]{br07} Branduardi-Raymont, G., Bhardwaj, A., Elsner, R.F., Gladstone, G.R., Ramsay, G., Rodriguez, P., Soria, R., Waite, J.H.,Jr., Cravens, T.E. 2007 A\&A 463, 761
\bibitem[Chandra et al.(2015)]{cha15} Chandra, P., Wade, G.A., Sundqvist, J.O. et al. 2015 MNRAS 452, 1245 
\bibitem[Chen \& White(1991)]{cw91}  Chen, W., White, R.L. 1991 ApJ 381, 63
\bibitem[Chlebowski(1989)]{chl89} Chlebowski, T. 1989 ApJ 342, 1091 
\bibitem[de Menezes et al.(2021)]{men21} de Menezes, R., Orlando, E., Di Mauro, M., Strong, A. 2021 in press, DOI:10.1093/mnras/stab2150
\bibitem[Donati et a.(2002)]{don02} Donati, J.-F., Babel, J., Harries, T.J., Howarth, I.D., Petit, P., Semel, M. 2002 MNRAS 333, 55
\bibitem[Eichler \& Usov(1993)]{eu93} Eichler, D., \& Usov, V. 1993 ApJ, 402, 271
\bibitem[Farnier et al.(2010)]{far10} Farnier, C., Walter, R., Leyder, J.-C. 2010 A\&A 526, A57
\bibitem[Funk et al.(2013)]{fun13} Funk, S., Hinton, J.A. (CTA Consortium) 2013, APh 43, 348
\bibitem[Gagne et al.(2005)]{gag05} Gagn\'e, M., Oksala, M.E., Cohen, D.H., Tonnesen, S.K., ud-Doula, A., Owocki, S.P., Townsend, R.H.D., MacFarlane, J.J. 2005 ApJ 628, 986 
\bibitem[Kraus et al.(2009)]{kra09} Kraus, S., Weigelt, G., Balega, Y.Y. et al. 2009 A\&A 497, 195
\bibitem[Leto et al(2006)]{le06} Leto, P., Trigilio, C., Buemi, C.S., Umana, G., Leone, F. 
2006 A\&A 458, 831
\bibitem[Leto et al(2017)]{le17} Leto, P., Trigilio, C., Oskinova, L. et al. 2017 MNRAS 457, 2820
\bibitem[Linsky et al.(1992)]{lin92} Linsky, J.L., Drake, S.A., Bastian, T. S. 1992 ApJ 393, 341
\bibitem[Lucy(1980)]{luc80} Lucy, L.B. 1982, ApJ, 255, 286
\bibitem[Lucy \& White(1982)]{luc82} Lucy, L.B., \& White, R.L. 1980 ApJ 241, 300
\bibitem[Maier et al.(20170]{mai17} Maier, G. et al. (CTA Consortium) 2017, in Proc. 35th ICRC 
(Busan, Korea), PoS(ICRC2017), 846
\bibitem[Marchili et al.(2018)]{mar18} Marchili, N., Piano, G., Cardillo, M., et al. 2018 A\& A, 615, A82
\bibitem[Moskalenko et al.(2006)]{mos06} Moskalenko, I. V., Porter, T. A., \& Digel, S. W. 2006
ApJ 652, L65
\bibitem[Orlando \& Strong(2007)]{os07} Orlando, E., \& Strong, A. W. 2007 Ap\& SS 309, 359
\bibitem[Orlando \& Strong(2008)]{os08} Orlando, E., \& Strong, A. W. 2008 A\& A 480, 84
\bibitem[Orlando \& Strong(2021)]{os21} Orlando, E., \& Strong, A. W. 2021 JCAP 04, 004
\bibitem[Owocki \& Rybicki(1984)]{or84} Owocki, S.P., Rybicki,G.B. 1984,  ApJ 284, 337
\bibitem[Petit et al.(2013)]{pet13} Petit,V., Owocki, S.P., Wade, G.A. et al. 2013 MNRAS 429, 398
\bibitem[Pittard \& Dougherty(2006)]{pd06} Pittard, J. M. \& Dougherty, S.M. 2006 MNRAS 372, 801
\bibitem[Pollock(1987)]{pol87} Pollock, A.M.T. 1987 ApJ 320, 283
\bibitem[Pshirkov et al.(2016)]{psh16} Pshirkov, M. S. 2016 MNRAS 457, L99
\bibitem[Reimer et al.(2006)]{rpr06} Reimer, A., Pohl, M., Reimer, O. 2006 ApJ 644, 1118
\bibitem[Revinius et al(2013)]{rev13} Revinius, T., Townsend R.H., Kochukhov O. et al. 2013 
MNRAS 429, 177
\bibitem[Shore(1987)]{sho87} Shore, S.N. 1987 AJ 94, 731
\bibitem[Shore \& Brown(1990)]{sb90} Shore, S.N., Brown, D.N. 1990 ApJ 365, 665 
\bibitem[Sierpowska \& Bednarek(2005)]{sb05} Sierpowska, A., Bednarek, W. 2005 MNRAS 356, 711
\bibitem[Sim\'on-Diaz et al.(2006)]{sim06} Sim\'on-Diaz, S., Herrero, A., Esteban, C., 
Najarro, F.  2006 A\&A 448, 351
\bibitem[Stahl et al.(1993)]{sta93} Stahl, O., Wolf, B., Gang, Th., Gummersbach, C.A., Kaufer, A., Kovacs, J., Mandel, H., Szeifert, Th. 1993 A\&A 274, L29
\bibitem[Stahl et al.(1996)]{sta96} Stahl, O., Kaufer, A., Rivinius, T. et al. 1996 A\&A 312, 539 
\bibitem[Tam et al.(2011)]{tam11} Tam, P.H.T., Huang, R.H.H., Takata, J., Hui, C.Y., Kong, A.K.H., Cheng, K.S. 2011 ApJ 736, L10
\bibitem[Tavani et al.(2009)]{tav09} Tavani, M., Sabatini, S., Pian, E. et al. 2009 ApJ 698 L142
\bibitem[Trigilio et al.(2004)]{tri04} Trigilio, C., Leto, P., Umana, G., Leone, F., Buemi, C.S. 
2004 A\&A 418, 593
\bibitem[ud-Doula \& Owocki(2002)]{uo02} ud-Doula, A., Owocki, S.P. 2002 ApJ 576, 413
\bibitem[Usov \& Melrose(1992)]{um92} Usov, V.V., Melrose D.B. 1992 ApJ 395, 575
\bibitem[Wade et al. 2006]{wad06} Wade, G.A., Fukkerton, A.W., Donati, J.-F., Landstreet, J.D., 
Petit, P., Strasser, S. 2006 A\&A 451, 195
\bibitem[White \& Chen(1992)]{wc92} White, W., \& Chen, W. 1992 ApJ 387, 81
\end{thebibliography}
\end{document}